\documentclass[acmsmall]{acmart}
\AtBeginDocument{%
  }

\usepackage[vlined, linesnumbered, ruled]{algorithm2e}

\SetKw{KwBy}{with step}
\SetCommentSty{commentfont}
\SetKwComment{AlgoComment}{// }{}

\usepackage{tikz}
\usepackage{amsthm}
\usepackage{hyperref}
\usepackage{cleveref}
\usepackage{tikz}
\usepackage{booktabs}
\usepackage{bm}
\usepackage{ifthen}
\usepackage{caption}
\usepackage{fancyhdr}
\usetikzlibrary{
    matrix,
    positioning,
    decorations.markings,
    decorations.pathreplacing,
    decorations.pathmorphing,
    angles,
    quotes,
    shadings,
    calc,
    patterns,
    backgrounds,
    fit,
    arrows,
    arrows.meta,
    shapes}
\usepackage{xcolor}
\usepackage{pgfkeys}
\usepackage[]{todonotes}
\usepackage{csquotes}
\usepackage[outline]{contour}
\usepackage{wrapfig}

\usepackage[inline]{enumitem} %

\usepackage{ifthen}
\newtoggle{extended}

\newtoggle{includeappendix}

\definecolor{DBRgreen}{RGB}{17, 87, 64}%
\definecolor{DBRblack}{RGB}{0, 0, 0}%
\definecolor{DBRwhite}{RGB}{255, 255, 255}%
\definecolor{DBRgray}{RGB}{156, 156, 156}%
\definecolor{DBRred}{RGB}{166, 25, 46}%
\definecolor{DBRblue}{RGB}{0, 45, 114}%
\definecolor{DBRorange}{RGB}{255, 102, 0}%
\definecolor{DBRbrown}{RGB}{148, 89, 24}%
\definecolor{DBRgrape}{RGB}{111, 45, 168}%
\definecolor{DBRgold}{RGB}{212, 175, 55}%

\definecolor{amber}{rgb}{1.0, 0.75, 0.0}
\definecolor{brightlavender}{rgb}{0.75, 0.58, 0.89}
\definecolor{airforceblue}{rgb}{0.36, 0.54, 0.66}
\definecolor{amethyst}{rgb}{0.6, 0.4, 0.8}
\definecolor{azure}{rgb}{0.0, 0.5, 1.0}
\definecolor{brightpink}{rgb}{1.0, 0.0, 0.5}
\definecolor{burgundy}{rgb}{0.5, 0.0, 0.13}
\definecolor{cornflowerblue}{rgb}{0.39, 0.58, 0.93}
\definecolor{darkchestnut}{rgb}{0.6, 0.41, 0.38}
\definecolor{emerald}{rgb}{0.31, 0.78, 0.47}
\definecolor{electricblue}{rgb}{0.49, 0.98, 1.0}

\colorlet{DBRFgreen}{DBRgreen!50!white}
\colorlet{DBRFred}{DBRred!50!white}
\colorlet{DBRFblue}{DBRblue!50!white}
\colorlet{DBRFgrape}{DBRgrape!50!white}
\colorlet{DBRgoldFgrape}{DBRgold!50!white}

\ifdefined\theorem
\else
        \newtheorem{theorem}{Theorem}
        \newtheorem{lemma}[theorem]{Lemma}
        \newtheorem{corollary}[theorem]{Corollary}
\fi

\newenvironment{cleverproof}[1]
{\begin{proof}[Proof of \Cref{#1}]}
                {\end{proof}}

\DeclareMathOperator{\simjoin}{\widetilde{\bowtie}}
\newcommand{\pc}[1]{\ensuremath{\mathtt{POPCNT}\!\left(#1\right)}}
\newcommand{\prob}{\ensuremath{\mathbb{P}}}

\pgfdeclarehorizontalshading{blueorange}{1cm}{%
  color(0)=(DBRblue);
  color(1)=(DBRorange)
}

\fancypagestyle{arxivnotice}{%
  \fancyhf{}%
  \fancyhead[C]{%
    \small\sffamily
    Accepted for publication in \textbf{Proc. ACM Manag. Data (PACMMOD / SIGMOD 2027)}.\\
    Please cite the official version: \url{https://doi.org/10.1145/3856354}.%
  }%
}

\begin{document}
\title{Joining Large String Collections with Jaro-Winkler Similarity}

\author{Manuel Lankes}
\correspondingauthor
\orcid{0009-0006-9769-0109}
\affiliation{%
    \institution{University of Salzburg}
    \streetaddress{Jakob-Haringer-Strasse 2}
    \city{Salzburg}
    \country{Austria}
    \postcode{5020}
}
\email{manuel.lankes@plus.ac.at}

\author{Daniel Kocher}
\orcid{0009-0003-3742-5555}
\affiliation{%
    \institution{University of Salzburg}
    \streetaddress{Jakob-Haringer-Strasse 2}
    \city{Salzburg}
    \country{Austria}
    \postcode{5020}
}
\email{dkocher@cs.sbg.ac.at}

\author{Daniel Schmitt}
\orcid{0009-0005-7656-7526}
\affiliation{%
    \institution{University of Salzburg}
    \streetaddress{Jakob-Haringer-Strasse 2}
    \city{Salzburg}
    \country{Austria}
    \postcode{5020}
}
\email{danielulrich.schmitt@plus.ac.at}

\author{Willi Mann}
\orcid{0000-0002-6460-6306}
\affiliation{%
    \institution{Celonis}
    \streetaddress{Theresienstrasse 6}
    \city{Munich}
    \country{Germany}
    \postcode{80333}
}
\email{w.mann@celonis.com}

\author{Thomas Samy Dafir}
\orcid{0009-0008-0153-7762}
\affiliation{%
    \institution{Salzburger Festspielfonds}
    \streetaddress{Hofstallgasse 1}
    \city{Salzburg}
    \country{Austria}
    \postcode{5020}
}
\authornote{Work carried out while at University of Salzburg.}
\email{samydafir@protonmail.com}

\author{Bin Shen}
\orcid{0009-0007-3004-015X}
\affiliation{%
    \institution{Celonis}
    \city{Redwood City}
    \state{CA}
    \country{USA}
}
\email{b.shen@celonis.com}

\author{Yujian Xu}
\orcid{0009-0000-5046-8523}
\affiliation{%
    \institution{Nikeonis Inc.}
    \streetaddress{2099 First Street SW, Suite 2250}
    \city{Washington}
    \state{DC}
    \postcode{20024}
    \country{USA}
}
\authornote{Work carried out while at Celonis.}
\email{yujian@nikeonis.com}

\author{Nikolaus Augsten}
\orcid{0000-0002-3036-6201}
\affiliation{%
    \institution{University of Salzburg}
    \streetaddress{Jakob-Haringer-Strasse 2}
    \city{Salzburg}
    \country{Austria}
    \postcode{5020}
}
\email{nikolaus.augsten@plus.ac.at}

\renewcommand{\shortauthors}{Lankes et al.}

\newcommand{\jaroWinkler}{S_W}
\newcommand{\jaro}{S_J}
\newcommand{\winklerTau}{\tau_W}
\newcommand{\jaroTau}{\tau_J}
\newcommand{\bitmap}{\ensuremath{F}}
\newcommand{\bitmaprec}[2]{\ensuremath{\bitmap_{#1}^{#2}}}  %
\newcommand{\bitmaprecsimple}[1]{\ensuremath{\bitmap_{#1}}}   %
\newcommand{\vecval}{\ensuremath{V}}
\newcommand{\vecrec}[2]{\ensuremath{\vecval_{#1}^{#2}}}  %
\newcommand{\vecrecsimple}[1]{\ensuremath{\vecval_{#1}}}   %
\newcommand{\heuristicvalue}{\ensuremath{H}}
\newcommand{\heuristic}[2]{\ensuremath{\heuristicvalue_{#1}^{#2}}}  %
\newcommand{\heuristicsimple}[1]{\ensuremath{\heuristicvalue_{#1}}}   %
\newcommand{\exactvalue}{\ensuremath{E}}
\newcommand{\exact}[2]{\ensuremath{\exactvalue_{#1}^{#2}}}  %
\newcommand{\exactsimple}[1]{\ensuremath{\exactvalue_{#1}}}   %
\newcommand{\TODO}[1]{} %
\newcommand{\Comment}[1]{\textcolor{blue}{#1}}
\newcommand{\unionbitm}{\ensuremath{\circ}} %

\newcommand{\setuniverse}{\ensuremath{\Gamma}} %
\newcommand{\alphabet}{\ensuremath{\Sigma}} %

\newcommand{\RS}{\ensuremath{{R}\text{-}{S}}} %

\newif\ifshowcolors

\ifshowcolors
  \colorlet{R1color}{DBRorange}
  \colorlet{R2color}{DBRblue!50!white}
  \colorlet{R3color}{DBRgrape}

  \colorlet{R12color}{brightlavender}
  \colorlet{R13color}{emerald}
  \colorlet{R23color}{darkchestnut}

  \colorlet{R123color}{DBRgoldFgrape}
  \colorlet{RelWorkcolor}{DBRFgreen!100!white}
\else
  \colorlet{R1color}{black}
  \colorlet{R2color}{black}
  \colorlet{R3color}{black}

  \colorlet{R12color}{black}
  \colorlet{R13color}{black}
  \colorlet{R23color}{black}

  \colorlet{R123color}{black}
  \colorlet{RelWorkcolor}{black}
\fi

\pgfkeys{
    /reviewer/1/.initial   = R1color,
    /reviewer/2/.initial   = R2color,
    /reviewer/3/.initial   = R3color,
    /reviewer/12/.initial  = R12color,
    /reviewer/21/.initial  = R12color,
    /reviewer/13/.initial  = R13color,
    /reviewer/31/.initial  = R13color,
    /reviewer/23/.initial  = R23color,
    /reviewer/32/.initial  = R23color,
    /reviewer/123/.initial = R123color,
    /reviewer/RelWork/.initial = RelWorkcolor,
    /reviewer/default/.initial = DBRblack
}

\newcommand{\review}[2][1]{%
    \textcolor{\pgfkeysvalueof{/reviewer/#1}}{#2}%
}

\newcommand{\relwork}[1]{%
    {\color{\pgfkeysvalueof{/reviewer/RelWork}}#1}%
}

\newcommand{\dressler}{\textsc{Dressler}\xspace}
\newcommand{\keil}{\textsc{Keil}\xspace}

\newcommand{\puma}{\textsc{Puma}\xspace}
\newcommand{\pumap}{\textsc{Puma+}\xspace}

\newcommand{\collin}{LIN}
\newcommand{\colxor}{XOR}
\newcommand{\colset}{SET}

\newcommand{\algname}[4]{\textsc{#1#2#3#4}\xspace}
\newcommand{\pumaplusstring}{+}

\newcommand{\sizefilterstring}{sf}
\newcommand{\sizefilter}{\textsc{\sizefilterstring}\xspace}

\newcommand{\prefixstring}{Pre}
\newcommand{\prefix}{\textsc{\prefixstring}\xspace}

\newcommand{\pallocstring}{Pal}
\newcommand{\palloc}{\textsc{\pallocstring}\xspace}

\newcommand{\dibstring}{Dib}
\newcommand{\dib}{\textsc{\dibstring}\xspace}

\newcommand{\mergestring}{Merge}
\newcommand{\merge}{\textsc{\mergestring}\xspace}

\newcommand{\sandbitstring}{Sand}
\newcommand{\sandbit}{\textsc{\sandbitstring}\xspace}

\newcommand{\sibtstring}{Sib}
\newcommand{\sibt}{\textsc{\sibtstring}\xspace}

\newcommand{\roaringstring}{Roar}
\newcommand{\roaring}{\textsc{\roaringstring}\xspace}

\newcommand{\fallbackstring}{Fallback}
\newcommand{\fallback}{\textsc{\fallbackstring}\xspace}

\newcommand{\linearverstring}{Jump}
\newcommand{\linearver}{\textsc{\linearverstring}\xspace}

\newcommand{\linearverstringnosort}{JumpNoSort}
\newcommand{\linearvernosort}{\textsc{\linearverstringnosort}\xspace}

\newcommand{\quadverstring}{Quad}
\newcommand{\quadver}{\textsc{\quadverstring}\xspace}

\newcommand{\rapidfuzzstring}{Rfuzz}
\newcommand{\rapidfuzz}{\textsc{\rapidfuzzstring}\xspace}

\newcommand{\BasakFlatString}{BasakOpt}
\newcommand{\basakflat}{\textsc{\BasakFlatString}\xspace}

\newcommand{\BasakUnoptString}{BasakList}
\newcommand{\basakunopt}{\textsc{\BasakUnoptString}\xspace}

\newcommand{\sfdibp}{\algname{\sizefilterstring}{\dibstring}{}{\pumaplusstring}}
\newcommand{\sfmergep}{\algname{\sizefilterstring}{\mergestring}{}{\pumaplusstring}}
\newcommand{\sfsandp}{\algname{\sizefilterstring}{\sandbitstring}{}{\pumaplusstring}}
\newcommand{\sfsibp}{\algname{\sizefilterstring}{\sibtstring}{}{\pumaplusstring}}
\newcommand{\sfroarp}{\algname{\sizefilterstring}{\roaringstring}{}{\pumaplusstring}}

\newcommand{\diblinver}{\algname{}{\dibstring}{\linearverstring}{}}
\newcommand{\sfdiblinver}{\algname{\sizefilterstring}{\dibstring}{\linearverstring}{}}
\newcommand{\sfdiblinverp}{\algname{\sizefilterstring}{\dibstring}{\linearverstring}{\pumaplusstring}}

\newcommand{\sfdibrfuzzrp}{\algname{\sizefilterstring}{\dibstring}{\rapidfuzzstring}{\pumaplusstring}}

\newcommand{\prediblinver}{\algname{\prefixstring}{\dibstring}{\linearverstring}{}}
\newcommand{\prediblinverp}{\algname{\prefixstring}{\dibstring}{\linearverstring}{\pumaplusstring}}

\newcommand{\paldiblinver}{\algname{\pallocstring}{\dibstring}{\linearverstring}{}}
\newcommand{\paldiblinverp}{\algname{\pallocstring}{\dibstring}{\linearverstring}{\pumaplusstring}}

\newcommand{\sfrfuzz}{\algname{\sizefilterstring}{\rapidfuzzstring}{}{}}
\newcommand{\sfdibrfuzzp}{\algname{\sizefilterstring}{\dibstring}{\rapidfuzzstring}{\pumaplusstring}}

\newcommand{\figscale}{1.}

\begin{abstract}
    The Jaro-Winkler measure, originally developed for U.S.\ Census record linkage, captures similarities between strings and accounts for common typographical variations. Despite its widespread use, finding all string pairs that exceed a given Jaro-Winkler threshold %
    remains challenging as the string collection grows large. 
    \relwork{Existing algorithms for computing the Jaro-Winkler similarity between two strings are either slow for short strings or require quadratic time and fail to scale to long strings.
    }
    To reduce the number of %
    string comparisons, state-of-the-art techniques \relwork{perform an index-nested-loop join on trie-based indexes}; however, they have only been evaluated on small datasets and high similarity thresholds. %
    For lower similarity thresholds that are relevant in practice, these techniques incur an excessive number of costly Jaro-Winkler computations.
    
    We present a novel approach for efficient Jaro-Winkler similarity joins on large string collections.
    Our solution is based on three key techniques:
    (1) \relwork{\emph{\linearver} is a fast algorithm for computing the Jaro-Winkler similarity between two strings that runs in linear time for bounded alphabets.}
    (2) \relwork{\emph{Dual-bound intersection bitmaps} encode strings} -- represented as character sets -- into fixed-size bitmaps, which are used to prune candidate string pairs by providing effective bounds on their character overlap; unlike prior bitmap techniques, our bitmaps are equipped with a cost model and adapt to the input data.
    (3) Our \emph{\puma framework} groups strings by prefix to tighten filter conditions, thereby reducing the number of similarity computations.
    Extensive experiments on large string collections demonstrate that our approach consistently outperforms the state of the art, achieving speedups of up to three orders of magnitude.
    
\end{abstract}

\ccsdesc[500]{Information systems~Data structures}
\ccsdesc[500]{Information systems~Data cleaning}
\ccsdesc[500]{Information systems~Proximity search}

\setcopyright{cc}
\setcctype{by}
\acmJournal{PACMMOD}
\acmYear{2026} \acmVolume{4} \acmNumber{6 (SIGMOD)} \acmArticle{359}
\acmMonth{12} \acmDOI{10.1145/3856354} 

\keywords{Similarity Joins, String Similarity, Jaro-Winkler, Bitmap Filters}

\received{17 April 2026}
\received[revised]{19 August 2026}
\received[accepted]{12 September 2026}

\maketitle
\thispagestyle{arxivnotice}

\section{Introduction} \label{sec:introduction}

Data is generated at an unprecedented rate, and the ability to process and correlate it becomes increasingly important. Thus, similarity joins have become a fundamental operation in database management systems. A similarity join between two collections of objects, $R$ and $S$, returns all pairs of objects $(r, s) \in R \times S$ whose similarity exceeds a user-defined threshold. Similarity joins have been extensively studied across various domains~\cite{Mann:EmpiricalEvaluationSetSimilarityJoin}, and prominent industry applications include Google’s near-duplicate web page detection~\cite{Henzinger:FindingNearDuplicateWebPages}, Celonis’ advanced process mining and invoice deduplication~\cite{Kocher:ScalingDensityBasedClustering, Celonis:AlgorithmAndDefaultSearchPatterns}, and Uber’s fraud detection system~\cite{Metwally:SimilarityJoinsSparseFeatures}.

In this paper, we study \emph{similarity joins on strings} based on the \emph{Jaro-Winkler similarity} measure: Given a user-defined threshold $\tau_W$, all pairs of strings that satisfy this threshold are computed. Jaro-Winkler~\cite{Jaro:AdvancesRecordLinkageMethodology,Winkler:StringComparatorMetricsEnhancedDecisionRules} was developed for record linkage on strings at the U.S. Census Bureau and is based on extensive practical experience in data matching~\cite{christen-data_matching-2012}. The similarity of two strings $r$ and $s$ is assessed in two steps:
(1) \emph{Jaro similarity:} Compute a similarity value based on the number of matching characters $m$ and character transpositions $t$.
(2) \emph{Winkler boost:} Increase the Jaro value based on the length $l$ of a common prefix within the first four characters of $r$ and $s$.

Its flexibility has made Jaro-Winkler popular for various matching tasks on short strings, including record linkage on names~\cite{Kloo:ImprovingRecordLinkage,Elmagarmid:DuplicateRecordDetectionSurvey}, data cleaning~\cite{Prokoshyna:CombiningQuantitativeAndLogicalDataCleaning}, and entity linking~\cite{Galarraga:CanonicalizingOpenKnowledgeBases}.
However, joining large string collections using Jaro-Winkler remains challenging.
Existing solutions, which employ trie-based indexes to reduce the number of string comparisons, have only been evaluated on small datasets with fewer than 10k strings~\cite{Dressler:OnEfficientExecutionBoundedJaroWinklerDistances}, or at high similarity thresholds $\winklerTau > 0.9$~\cite{Keil:EfficientBoundedJaroWinklerSimilarityBasedSearch}.
These approaches become ineffective at the lower thresholds relevant in practice, e.g., $\winklerTau = 0.85$~\cite{Kloo:ImprovingRecordLinkage}, and can take many hours even on medium-sized datasets of $10^5$ strings.

\begin{figure}[t]
    \centering
    \begin{tikzpicture}[semithick]
        \newcommand{\gear}[6]{%
            \foreach \i in {1,...,#1} {%
                    \begin{scope}[rotate=(\i-1)*360/#1]
                        \draw[#6,thick]
                        (0:#2) arc (0:#4:#2)
                        [rounded corners=1.5pt]
                        to (#4+#5:#3)
                        arc (#4+#5:360/#1-#5:#3)
                        to (360/#1:#2);
                    \end{scope}%
                }%
        }
        
        \tikzset{
            stage/.style={
                    minimum height=2.35cm,
                    inner sep=0pt,
                    outer sep=0pt
                },
            geartext/.style={draw=none},
            flow/.style={-stealth, ultra thick},
            inflow/.style={stealth-, ultra thick},
            sidelabel/.style={black!70},
        }
        \def\stagegap{2.2cm}
        \def\pumawidth{2cm}
        \def\dibwidth{2.1cm}
        \def\verwidth{1.5cm}
        \def\topgearx{0.75cm}
        \def\topgeary{-0.75cm}
        \def\pumageardx{0.45cm}
        \def\geardx{1.3em}
        \def\geardown{-1.95em}
        \def\flowy{-0.cm}
        \def\stageradius{4pt}
        \def\stringInY{0.425cm}
        \def\thresholdInY{-0.525cm}
        \newcommand{\drawstage}[1]{%
            \path[draw=black, fill=gray!5]
            (#1.north west) --
            ($(#1.north east)+(-\stageradius,0)$)
            arc[start angle=90,end angle=0,radius=\stageradius] --
            ($(#1.south east)+(0,\stageradius)$)
            arc[start angle=0,end angle=-90,radius=\stageradius] --
            ($(#1.south west)+(\stageradius,0)$)
            arc[start angle=-90,end angle=-180,radius=\stageradius] --
            (#1.north west);
        }
        
        \node[stage, minimum width=\pumawidth, anchor=north west] (pumaBox) at (-0.75cm,0.6cm) {};
        \node[stage, minimum width=\dibwidth, anchor=north west] (dibBox) at ([xshift=\stagegap]pumaBox.north east) {};
        \node[stage, minimum width=\verwidth, anchor=north west] (verBox) at ([xshift=\stagegap]dibBox.north east) {};
        \drawstage{pumaBox}
        \drawstage{dibBox}
        \drawstage{verBox}
        
        \coordinate (pumaPrefix) at ([xshift=\topgearx,yshift=\topgeary]pumaBox.north west);
        \coordinate (pumaJaro) at ([xshift=\pumageardx,yshift=\geardown]pumaPrefix);
        \coordinate (dibLowerBound) at ([xshift=4.05cm]pumaPrefix);
        \coordinate (dibUpperBound) at ([xshift=\geardx,yshift=\geardown]dibLowerBound);
        \coordinate (dibOverlap) at ([xshift=\geardx,yshift=0.69cm]dibUpperBound);
        \coordinate (verLinear) at ([yshift=-0.65cm]verBox.north);
        
        \begin{scope}[shift={(pumaPrefix)}, local bounding box=pumaPrefixGear]
            \gear{8}{0.3}{0.45}{20}{2}{red}
            \node[geartext, red] at (0,0) {\textbf{P}};
        \end{scope}
        
        \begin{scope}[shift={(pumaJaro)}, rotate=25, local bounding box=pumaJaroGear]
            \gear{8}{0.3}{0.45}{20}{2}{blue}
            \node[geartext, blue] at (0,0) {$\boldsymbol{\tau_J}$};
        \end{scope}
        
        \begin{scope}[shift={(dibLowerBound)}, local bounding box=dibLowerBoundGear]
            \gear{8}{0.3}{0.45}{20}{2}{green!75!black}
            \node[geartext, green!75!black] at (0,0) {$\boldsymbol{\varphi_l}$};
        \end{scope}
        
        \begin{scope}[shift={(dibUpperBound)}, rotate=25, local bounding box=dibUpperBoundGear]
            \gear{8}{0.3}{0.45}{20}{2}{black}
            \node[geartext, black] at (0,0) {$\boldsymbol{\varphi_u}$};
        \end{scope}
        
        \begin{scope}[shift={(dibOverlap)}, rotate=45, local bounding box=dibOverlapGear]
            \gear{8}{0.3}{0.45}{20}{2}{orange}
            \node[geartext, orange] at (0,0) {\textbf{O}};
        \end{scope}
        
        \begin{scope}[shift={(verLinear)}, yshift=-12.5,rotate=25, local bounding box=verLinearGear]
            \gear{8}{0.4}{0.55}{20}{2}{black}
            \node[geartext, black] at (0,0) {\footnotesize $\boldsymbol{\mathcal{O}(n)}$};
        \end{scope}
        
        \node[black] at ([yshift=0.2cm]pumaBox.south) {\textbf{\puma{}}};
        \node[black] at ([yshift=0.2cm]dibBox.south) {\textbf{\dib{}}};
        \node[black] at ([yshift=0.2cm]verBox.south) {\textbf{\linearver{}}};
        
        \coordinate (stringsIn) at ([yshift=\stringInY]pumaBox.west);
        \coordinate (thresholdIn) at ([yshift=\thresholdInY]pumaBox.west);
        \coordinate (pumaOut) at ([yshift=\flowy]pumaBox.east);
        \coordinate (dibOut) at ([yshift=\flowy]dibBox.east);
        
        \node[minimum width=2em, inner sep=0] (result) at ([xshift=0.975cm,yshift=-0.cm]verBox.east) {\raisebox{0.06cm}{$\boldsymbol{R\simjoin S}$}};
        
        \draw[inflow] (stringsIn) -- ++(-0.6cm,0) -- ++(0,0.cm);
        \draw[inflow] (thresholdIn) -- ++(-0.6cm,0) -- ++(0,-0.cm);
        \draw[flow] (pumaOut) -- ++(\stagegap,0)
        node[pos=0.5, anchor=south, sidelabel] {\rotatebox{0}{\scriptsize \textbf{pre-candidates}}};
        \draw[flow] (dibOut) -- ++(\stagegap,0)
        node[pos=0.5, anchor=south, sidelabel] {\rotatebox{0}{\scriptsize \textbf{candidates}}};
        \draw[flow] (verBox.east) -- (result.west);
        
        \node at ([xshift=-0.95cm,yshift=\stringInY]pumaBox.west) {$\boldsymbol{R}, \boldsymbol{S}$};
        \node at ([xshift=-0.9cm,yshift=\thresholdInY]pumaBox.west) {$\boldsymbol{\tau_W}$};
    \end{tikzpicture}
    \caption{Jaro-Winkler join between two string collections ${R}$, ${S}$ with threshold $\winklerTau$. %
        \puma{}: computes Jaro threshold $\tau_J$ based on prefix length P; \dib{}: performs overlap verification O based on lower/upper bounds ${\varphi_{l/u}}$; \linearver{}: linear-time Jaro-Winkler verification \relwork{for bounded alphabets.}}
    \label{fig:algo-outline}
    \Description{Conceptual flowchart of the three main components of our Jaro-Winkler similarity join algorithm. Inputs consist of two string collections R and S, and a threshold tau_W. Stage 1 is PUMA, which takes the inputs and computes pre-candidates using prefix length P and Jaro threshold tau_J. Stage 2 is DIB, which processes pre-candidates using lower and upper bounds phi_l and phi_u and an overlap filter to produce candidates. Stage 3 is JUMP, which performs linear-time verification to produce the final similarity join result.}
\end{figure}
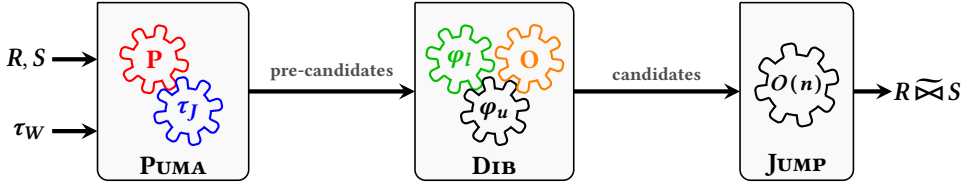

We identify three main challenges that make Jaro-Winkler joins difficult:
(1) Computing the Jaro similarity between two strings is expensive: Each character in one string must be compared to a window of linear size in the other string.
Therefore, \relwork{most} state-of-the-art algorithms run in quadratic time in the string length.
\relwork{While linear algorithms exist~\cite{Basak:OnComputingJaroSimilarity,Basak:FastAlgorithmsComputingJaroSimilarity}, optimized quadratic implementations~\cite{RapidFuzz:2025} consistently outperform them even for the longest strings in our evaluation (${>} 100$ characters).}
(2) Candidate filtering is challenging: Although -- like for Jaccard and Cosine similarity -- the character overlap of two strings provides a lower bound on the Jaro-Winkler similarity, this bound is hard to exploit.
The required overlap is lower than for other normalized similarity measures, which makes this bound less effective.
Furthermore, the number of distinct elements in the character sets is small, which renders common join techniques such as prefix filtering less effective~\cite{Schmitt:TwoLevelSignatureSchemeStableSetSimilarityJoinsTWOL}.
(3) Pairwise prefix dependency: The required character overlap that two strings must satisfy for a given Jaro-Winkler threshold depends on the length of their common prefix.
This inherent pairwise dependency makes it difficult to apply state-of-the-art set similarity join techniques. Conservatively assuming the maximum prefix length weakens the overlap bound and leads to even more candidates.
Figure~\ref{fig:algo-outline} shows how we address these challenges at three levels.

To handle prefix effects, we propose the \emph{Prefix-based Unified Multi-Join Aggregation (PUMA)} framework, which groups strings by common prefixes and uses this information to tighten filter conditions. To avoid duplicate candidates across groups, we introduce a traversal technique that eliminates the need for an expensive final deduplication step.

To generate candidates, we reduce the Jaro-Winkler join to a set overlap join.
We propose \emph{Dual-Bound Intersection Bitmaps (\dib{})}, which encode strings into fixed-size bitmaps to provide effective bounds on the character overlap:
Frequent characters map to dedicated bits to establish a lower bound, while the remaining characters are hashed to derive an upper bound, yielding an effective, low-overhead filter.
Compared to prior bitmap techniques for sets~\cite{Sandes:BitmapFilterSpeedingUpExactSetSimilarityJoinsBitwiseOperations}, \dib{}s provide both upper and lower bound, handle skewed character distributions, and adapt to the data via a cost model.

Finally, candidate pairs are verified using \linearver, \relwork{a new, \emph{merge-based Jaro-Winkler} algorithm.
\linearver avoids the quadratic window search of most prior methods, runs in linear time for bounded alphabets, and clearly outperforms all linear competitors in practice.}

Summarizing, our technical contributions are as follows:

\begin{itemize}
    \item Our \textbf{\puma framework} for Jaro-Winkler joins groups strings into subgroups that enable tighter filter conditions, and our traversal strategy avoids duplicate candidate pairs.
    \item Our fixed-size \textbf{dual-bound intersection bitmaps (\dib{}s)} provide an effective, low-overhead filter for set overlap, gracefully handle skewed set element frequencies, adapt to the data via a cost model, and apply beyond strings to set similarity queries in general.
    \item \relwork{Our verification algorithm, \textbf{\linearver}, computes the Jaro-Winkler similarity of two strings in linear time for bounded alphabets; existing approaches either require quadratic time in the string length or are slow in practice.}
    \item Experiments across a wide range of thresholds, including low similarities and large datasets with up to one million strings, show that our approach outperforms the state of the art, achieving speedups of up to three orders of magnitude.
\end{itemize}
\noindent

\paragraph{\textbf{Paper Outline}}
Section~\ref{sec:preliminaries} introduces the Jaro-Winkler similarity join problem.
Section~\ref{sec:algo-outline} provides an outline of our approach; the details are presented in Sections~\ref{sec:winkler-trie} (\puma), \ref{sec:overlap-join} (\dib), and \ref{sec:jw-join}~(\linearver), respectively.
\relwork{Section~\ref{sec:rs-joins} generalizes our approach to $\RS$ joins}, Section~\ref{sec:experiments} presents our experiments, Section~\ref{sec:related-work} discusses related work, and Section~\ref{sec:conclusion} concludes.

\section{Preliminaries} \label{sec:preliminaries}

In this section, we review the Jaro-Winkler similarity measure as well as two existing lower bounds from literature, and define the problem addressed in this paper.

\subsection{Jaro-Winkler Similarity} %

Jaro-Winkler~\cite{Winkler:StringComparatorMetricsEnhancedDecisionRules} measures the similarity between two strings $r$ and $s$.
It extends the \emph{Jaro similarity}~\cite{Jaro:AdvancesRecordLinkageMethodology,jaro-1978-original_Jaro_similarity} , which was introduced to match names in the U.S. Census.
A string $r$ is a sequence of $|r|$ symbols from a finite \emph{alphabet of symbols $\alphabet$}, $r[i]$ denotes the $i$-th character of $r$ (i.e., $r = r[1]r[2]\ldots r[|r|]$).
A character $r[i]$ in $r$ has a \emph{match} in $s$ if there is a character $s[j]$ such that $r[i] = s[j]$, and the positions $i$, $j$ of the two characters differ by at most $\lfloor \max\{|r|, |s|\} / 2 \rfloor - 1$.
Characters are matched from left (smaller position) to right; if a character $r[i]$ has multiple possible matching partners in $s$, it is matched to the first unmatched character in $s$.

\begin{definition}[Jaro Similarity]
    The Jaro similarity $\jaro: \alphabet^+ \times \alphabet^+ \to [0, 1]$ of two strings $r$, $s \in \alphabet^+$ is defined as $\jaro(r, s)=0$ if $m=0$, or
    \begin{align}
        \label{eq:jaro}
        \jaro(r, s) = \frac{1}{3} \left( \frac{m}{|r|} + \frac{m}{|s|} + \frac{m - t}{m} \right)\quad \text{for}\ m>0,
    \end{align}
    where $m$ is the number of matching characters between $r$ and $s$, and $t$ is the number of transpositions.
\end{definition}

The number of transposition is computed based on the \emph{matched substrings} $\hat{r}$ and $\hat{s}$, $|\hat{r}| = |\hat{s}|$, which are formed by removing all unmatched characters from $r$ and $s$.
The number of transpositions is $(\hat{r}\,\triangle\,\,\hat{s})/2$, where $\triangle$ denotes the Hamming distance, i.e., the number character positions $i\leq |\hat{r}|$ such that $\hat{r}[i]\neq \hat{s}[i]$.
As keypunch errors become more likely toward the end of a string, Winkler proposed boosting the similarity of strings with a common prefix~\cite{Winkler:OverviewRecordLinkage}.

\begin{definition}[Jaro-Winkler Similarity] \label{eq:jaro-winkler}
    \label{def:jaro-winkler-sim}
    Given a scaling factor $p$ and two strings $r$, $s$ with a common prefix length $l$, the \emph{Jaro-Winkler similarity} $\jaroWinkler: \alphabet^+ \times \alphabet^+ \to [0, 1]$ is defined as:
    \begin{align}
        \jaroWinkler(r, s) =
        \begin{cases}
            \jaro(r, s) + l \cdot p \cdot (1 - \jaro(r, s)), & \text{if } \jaro(r, s) \geq 0.7, \\
            \jaro(r, s),                                     & \text{otherwise.}
        \end{cases}
    \end{align}
\end{definition}

The common prefix length is limited to some $l_{\max}$ and the scaling factor controls the weight of the common prefix.
For the remainder of this work, we use the recommended standard setting for the Jaro-Winkler similarity, i.e., $l_{\max} = 4$ and $p = 0.1$~\cite{Winkler:OverviewRecordLinkage}.
We will develop the first declarative definition of a Jaro mapping in Section~\ref{sec:jaro-winkler-similarity-computation}.

\subsection{Problem Definition \& Objective} \label{sec:problem-definition}

Given two collections of strings $R$ and $S$, a similarity measure $\delta$, and a threshold $\tau \in [0, 1]$, a string similarity join $R\simjoin S$ finds all pairs of strings $r\in R$, $s\in S$ that exceed a given threshold $\tau$ w.r.t.\ $\delta$.
Specifically, we focus on the Jaro-Winkler (string) similarity join:
\begin{equation*}
    T = R\;\widetilde{\bowtie}\;S=\{(r, s) \in R \times S \mid \jaroWinkler(r, s) \geq \tau\}.
\end{equation*}
\review[23]{For simplicity, we focus on self-joins ($R=S$); $\RS$ joins ($R\neq S$) are discussed in Section~\ref{sec:rs-joins}.}

\emph{\textbf{Objective.}} Our goal is to develop an efficient algorithm for Jaro-Winkler similarity joins that scales to large string collections for a wide range of Jaro-Winkler thresholds.

\subsection{Jaro Threshold and Filters}

Given a Jaro-Winkler similarity threshold $\winklerTau$, we can derive a Jaro threshold $\jaroTau$ such that for any two strings $r$ and $s$ with a common prefix of length $l$:  $\jaroWinkler(r, s) \geq \winklerTau$ if and only if $\jaro(r, s) \geq \jaroTau$.
Based on Definition~\ref{def:jaro-winkler-sim}, we obtain the equivalent Jaro threshold $\jaroTau$ as
\begin{equation} \label{eq:jaro-threshold}
    \jaroTau = \max
    \begin{cases}
        \frac{\winklerTau - l \cdot p}{1 - l \cdot p}, \\
        \min\{0.7, \winklerTau\}.
    \end{cases}
\end{equation}

\noindent
This formulation is based on three key properties that collectively ensure $\jaroWinkler(r,s)\ge\winklerTau$:
(1) $\jaroTau \leq \winklerTau$ for all \(\winklerTau\in[0,1]\), \(l\in[0,4]\), and \(p=0.1\).
(2) If $\winklerTau<0.7$, then $\jaroTau=\winklerTau$.
(3) If $\winklerTau \geq 0.7$, but for a prefix length $l$ the fraction $\frac{\winklerTau - l \cdot p}{1 - l \cdot p}$ evaluates to a value below 0.7, then $\jaroTau$ is capped at $0.7$. 
This cap is required because a Jaro similarity $\jaro(r,s) < 0.7$ cannot be boosted by the prefix term to meet $\winklerTau \ge 0.7$.
Given a Jaro–Winkler threshold $\winklerTau$ and a common prefix of length $l$, $\smash{\jaroTau^l}$ denotes the equivalent Jaro threshold (cf.\ Equation~\ref{eq:jaro-threshold}).

\subsection{Jaro Filters}
Most similarity join algorithms are based on a filter-and-verify framework.
Bounds on the similarity measure are used to prune the search space and compute a set of candidates.
\relwork{Then, the similarity score is computed for each candidate to discard false positives and derive the result.}
\relwork{In the following, we discuss two filters for a given pair of strings $r$, $s$ and Jaro threshold $\jaroTau$.}

\paragraph{\textbf{Overlap Filter}}

The \textit{overlap filter}~\cite{Dressler:OnEfficientExecutionBoundedJaroWinklerDistances} bounds the number of common characters between $r$ and $s$. A pair $(r,s)$ that does not pass the overlap filter cannot satisfy threshold $\jaroTau$. We use \emph{counting tokenization}~\cite{Augsten:SimilarityJoinsRelationalDatabaseSystems} to transform strings into sets: Each character $a\in \alphabet$ in a string is represented by a unique token $a_i$ with $i$ being the occurrence count of $a$ in that string.
$\setuniverse$ denotes the \emph{universe} of all unique tokens over all obtained sets; note the subtle difference to the \emph{alphabet $\alphabet$} of distinct string characters.
For instance, the string $r = aabbcc$ is represented as the set $\{a_1, a_2, b_1, b_2, c_1, c_2\}$.
\relwork{The intersection of the token sets of $r$ and $s$ is $r \cap s\subseteq \setuniverse$, and the number of matched characters $m$ is bounded by this intersection: $m \leq |r \cap s|$.}
Based on Equation~\ref{eq:jaro}, we derive the overlap filter. Since transpositions reduce the Jaro similarity, we conservatively assume $t = 0$.

\begin{lemma}[Overlap Filter]
    \label{lem:overlap-filter}
    Given two strings, $r$ and $s$, and Jaro threshold $\jaroTau$:
    \[
        \jaro(r,s)\geq \jaroTau \Rightarrow  |r \cap s| \geq (3 \jaroTau - 1) \cdot \frac{|r| \cdot |s|}{|r| + |s|}.
    \]
\end{lemma}

\paragraph{\textbf{Size Filter}}

The \textit{size filter}~\cite{Dressler:OnEfficientExecutionBoundedJaroWinklerDistances} bounds the maximum length difference of two strings $r$ and $s$ that can satisfy a given Jaro threshold $\jaroTau$.
The maximum number of matched characters is bounded by the length of the shorter string: $m \leq \min\{|r|, |s|\}$.
Again, we assume $t = 0$ transpositions and derive the size filter from Equation~\ref{eq:jaro}.

\begin{lemma}[Size Filter] \label{lem:size-filter}
    Given two strings, $r$ and $s$, and Jaro threshold $\jaroTau> \frac{2}{3}$:
    \[
        \jaro(r,s)\geq \jaroTau> \frac{2}{3} \Rightarrow (3 \jaroTau - 2) \cdot |s| \leq |r| \leq |s| / (3 \jaroTau - 2).
    \]
\end{lemma}

\section{Algorithm Outline} \label{sec:algo-outline}

Our solution is composed of the three main components illustrated in Figure~\ref{fig:algo-outline}. 
\text{(1)~\textbf{\puma}} groups strings into subgroups that enable tighter Jaro similarity thresholds $\jaroTau$ for index-based filtering, thereby reducing the number of comparisons.
The output, called \emph{pre-candidates}, is the set of string pairs that pass the respective filter and must be considered under a specific Jaro threshold.
\textbf{(2)~\dib{}s:} Our dual-bound intersection bitmaps provide both lower and upper bounds on the overlap, enabling efficient overlap verification.
This step refines pre-candidates to \emph{candidates}, the subset that passes the overlap filter.
\textbf{(3)~\linearver{}:} All candidate string pairs are verified against the Jaro-Winkler threshold $\winklerTau$ \relwork{using our novel Jaro-Winkler verification, which runs in linear time for bounded alphabets}.

\section{The \puma Framework} \label{sec:winkler-trie}
Existing Jaro-Winkler joins perform an index-nested loop join on trie-based index structures to exploit common prefixes for batch filtering~\cite{Dressler:TimeEfficientExecutionBoundedJaroWinklerDistances,Dressler:OnEfficientExecutionBoundedJaroWinklerDistances} or verification~\cite{Keil:EfficientBoundedJaroWinklerSimilarityBasedSearch}.

In contrast, we reduce the Jaro-Winkler join to a set overlap join on tokenized strings.
Such reductions are common in set similarity joins for normalized measures such as Jaccard or Cosine~\cite{Mann:EmpiricalEvaluationSetSimilarityJoin}. Compared to these measures, Jaro-Winkler poses two challenges:
\begin{enumerate*}
    \item The required overlap depends not only on the token sets but also on the length of the common prefix, making standard set similarity joins not readily applicable.
    \item The required overlap is much smaller than for Jaccard or Cosine, reducing the effectiveness of filtering techniques.
\end{enumerate*}
As an example, consider two strings $r$ and $s$ of length $8$ and a typical threshold $\winklerTau = 0.85$. The required overlap is $|r \cap s| = 7$ with no common prefix, but only $|r \cap s| = 5$ for the maximum prefix length $l_{\max} = 4$, corresponding to a Jaccard threshold as low as $0.46$; in the literature, Jaccard thresholds are often $\geq 0.8$~\cite{Arasu:EfficientExactSetSimilarityJoins,Deng:EfficientPartitionBasedMethodExactSetSimilarityJoinsPARTALLOC,Xiao:EfficientSimilarityJoinsNearDuplicateDetectionPOSITIONAL}.

To handle prefix effects, we propose the \emph{Prefix-based Unified Multi-Join Aggregation (\puma)} framework, which groups strings by common prefix length $l$ to tighten filter conditions. For a given Jaro-Winkler threshold $\winklerTau$, each  length group has a Jaro threshold $\jaroTau^l$ (Equation~\ref{eq:jaro-threshold}), which determines the required overlap (cf.\ Lemma~\ref{lem:overlap-filter}). 

\paragraph{\textbf{Construction}} \label{sec:winkler-trie-construction}

\puma{} organizes strings in a prefix trie, inserting each string character by character.
An edge from depth $k$ to $k{+}1$ represents the $(k{+}1)$\textsuperscript{th} character.
Thus, the root contains all strings, and a node at depth $k$ contains all strings sharing the corresponding $k$-prefix.
Assuming $l_{\max} = 4$ for Jaro-Winkler, the trie has maximum depth $4$ and can be constructed in $\mathcal{O}(n)$ time for $n$ strings.

\paragraph{\textbf{Traversal}} \label{sec:winkler-trie-traversal}

The Jaro-Winkler join traverses all nodes of the trie. Let $n_k$ denote a node at depth $k$, $0 \leq k \leq l_{\max} = 4$, and $R_k$ the set of strings sharing the $k$-prefix of node $n_k$. For each node, we compute a partial Jaro join $T_k = R_k \simjoin R_k$ with threshold $\jaroTau^k$ (cf.\ Equation~\ref{eq:jaro-threshold} with $l = k$); note that multiple nodes may exist at depth $k > 0$. The overall result is the union of all partial joins. Since the partial joins are independent, the trie can be traversed in any order.

\paragraph{\textbf{Deduplication}} \label{sec:winkler-trie-duplicates}

Since string pairs $(r,s)$ with a long prefix also share shorter prefixes, each pair appears in up to $l_{\max}+1$ nodes, resulting in duplicates. This is undesirable as 
\begin{enumerate*}
    \item the same pair is processed in multiple nodes, and
    \item an expensive duplicate elimination step is required to derive the final result.
\end{enumerate*}
We address both issues by ensuring that each pair is processed exactly once: at the deepest node that stores both strings, i.e., the node corresponding to their longest common prefix.

Let $n_i$, $0 \leq i \leq l$, denote the nodes at depth $i$ that contain both $r$ and $s$. 
We distinguish two cases for processing $(r,s)$ at node $n_i$:
\begin{enumerate*}
    \item \emph{Computation:} We process $(r,s)$ at node $n_i$ using Jaro threshold $\jaroTau^i$ if $r$ and $s$ share a maximum prefix of length $i = l$ (i.e., $r[i+1] \neq s[i+1]$), or if $i = l_{\max}$.
    \item \emph{Pushdown:} Otherwise, the pair $(r,s)$ satisfies the \emph{pushdown condition} ($i < l_{\max}$ and $r[i+1] = s[i+1]$), and its processing is deferred to node $n_{i+1}$.
\end{enumerate*}
The pushdown is safe: if $(r,s)$ fails to meet $\smash{\jaroTau^{i+1}}$ at $n_{i+1}$, it also fails to meet the stricter threshold $\smash{\jaroTau^i}>\smash{\jaroTau^{i+1}}$ at $n_i$ (cf.\ Equation~\ref{eq:jaro-threshold}).

Computing the pre-candidates for each node $n_i$ is an orthogonal problem, and optimized join algorithms w.r.t.\ $\smash{\jaroTau^i}$ and the string set in the node may be used for this step. In our implementation, we use one of three alternatives: size filtering, a prefix index, or a partition-based index. The pushdown evaluation acts as a direct filter on pairs generated by the index. After probing the index with a query string, we evaluate the pushdown condition for each retrieved string and discard pairs that do not satisfy the condition.

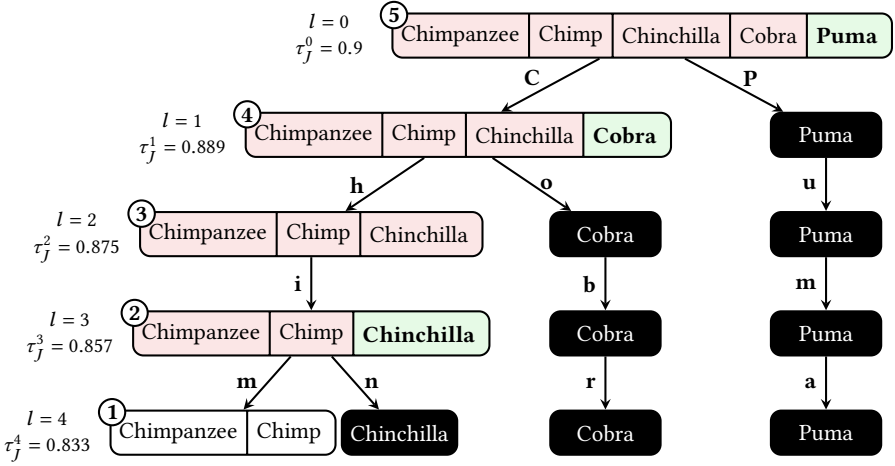
\begin{figure}[t]
     \centering
    \resizebox{0.9\linewidth}{!}{%
        \colorlet{group1}{red!85!black}
        \colorlet{group2}{green!80!black}
        \colorlet{singleEntry}{black}
        \colorlet{prunedEntry}{DBRblue!50!DBRwhite}
        \centering
        \tikzset{
            trie_node/.style={
                    draw,
                    rectangle,
                    rounded corners,
                    minimum width=1.5cm,
                    minimum height=0.6cm,
                    text opacity=1,
                    align=center,
                    font=\small
                },
            trie_node_split/.style={
                    trie_node,
                    rectangle split,
                    rectangle split horizontal,
                    rectangle split draw splits=true
                },
            trie_node5/.style={trie_node_split,rectangle split parts=5},
            trie_node4/.style={trie_node_split,rectangle split parts=4},
            trie_node3/.style={trie_node_split,rectangle split parts=3},
            trie_node2/.style={trie_node_split,rectangle split parts=2},
            trie_edge/.style={draw, -stealth},
            level_annotation/.style={font=\small, text=DBRblack},
            edge_label/.style={font=\small, text=DBRblack,left},
            nodecount/.style={
                    draw,
                    circle,
                    fill=white,
                    inner sep=1pt,
                    minimum size=1em,
                    font=\small
                },
        }
        \resizebox{\columnwidth}{!}{
            \begin{tikzpicture}[
                    thick,
                    -stealth,
                    level/.style={sibling distance=6.cm/#1,level distance=1.35cm},
                    level 1/.style={sibling distance=5.cm},
                    level 2/.style={sibling distance=4.cm},
                    level 4/.style={sibling distance=2.4cm}]
                \node [trie_node5,rectangle split part fill={group1,group1,group1,group1,group2},fill opacity=0.1] (Root) {
                    \nodepart{one}Chimpanzee
                    \nodepart{two}Chimp
                    \nodepart{three}Chinchilla
                    \nodepart{four}Cobra
                    \nodepart{five}\textbf{Puma}
                } child [trie_edge] {
                        node [trie_node4,rectangle split part fill={group1,group1,group1,group2},fill opacity=0.1] (N1) {
                                \nodepart{one}Chimpanzee
                                \nodepart{two}Chimp
                                \nodepart{three}Chinchilla
                                \nodepart{four}\textbf{Cobra}
                            }
                        child [trie_edge] {
                                node [trie_node3,rectangle split part fill={group1},fill opacity=0.1] (N2) {
                                        \nodepart{one}{Chimpanzee}
                                        \nodepart{two}{Chimp}
                                        \nodepart{three}{Chinchilla}
                                    }
                                child [trie_edge] {
                                        node [trie_node3,rectangle split part fill={group1, group1, group2},fill opacity=0.1] (N3) {
                                                \nodepart{one}Chimpanzee
                                                \nodepart{two}Chimp
                                                \nodepart{three}\textbf{Chinchilla}
                                            }
                                        child [trie_edge] {
                                                node [trie_node2, rectangle split part fill={white,white} ,fill opacity=0.1] (N4) {
                                                        \nodepart{one}{Chimpanzee}
                                                        \nodepart{two}{Chimp}
                                                    }
                                                edge from parent node[edge_label, yshift=0.cm] {\textbf{m}}
                                            }
                                        child [trie_edge] {
                                                node [trie_node,draw=singleEntry,fill=singleEntry,text=white] {Chinchilla}
                                                edge from parent node[edge_label, yshift=0.cm, xshift=0.45cm] {\textbf{n}}
                                            }
                                        edge from parent node[edge_label] {\textbf{i}}
                                    }
                                edge from parent node[edge_label, xshift=-0.15cm] {\textbf{h}}
                            }
                        child [trie_edge] {
                                node [trie_node,draw=singleEntry,fill=singleEntry,text=white] {Cobra}
                                child [trie_edge] {
                                        node [trie_node,draw=singleEntry,fill=singleEntry,text=white] {Cobra}
                                        child [trie_edge] {
                                                node [trie_node,draw=singleEntry,fill=singleEntry,text=white] {Cobra}
                                                edge from parent node[edge_label] {\textbf{r}}
                                            }
                                        edge from parent node[edge_label] {\textbf{b}}
                                    }
                                edge from parent node[edge_label, xshift=0.42cm] {\textbf{o}}
                            }
                        edge from parent node[edge_label, yshift=0.1cm] {\textbf{C}}
                    }
                child [trie_edge] {
                        node [trie_node,draw=singleEntry,fill=singleEntry,text=white] {Puma}
                        child [trie_edge] {
                                node [trie_node,draw=singleEntry,fill=singleEntry,text=white] {Puma}
                                child [trie_edge] {
                                        node [trie_node,draw=singleEntry,fill=singleEntry,text=white] {Puma}
                                        child [trie_edge] {
                                                node [trie_node,draw=singleEntry,fill=singleEntry,text=white] {Puma}
                                                edge from parent node[edge_label] {\textbf{a}}
                                            }
                                        edge from parent node[edge_label] {\textbf{m}}
                                    }
                                edge from parent node[edge_label] {\textbf{u}}
                            }
                        edge from parent node[edge_label, yshift=0.1cm, xshift=0.45cm] {\textbf{P}}
                    };

                \foreach \nodeName/\count/\l/\t in {Root/5/0/0.9, N1/4/1/0.889, N2/3/2/0.875, N3/2/3/0.857, N4/1/4/0.833} {
                        \node[draw=none] at ($(\nodeName.north west)!0.35!(\nodeName.west)$) (phNorth\nodeName) {};
                        \node[draw=none] at ($(\nodeName.south west)!0.35!(\nodeName.west)$) (phSouth\nodeName) {};
                        \node[draw=none,rounded corners,minimum width=4.1em] at ($(phNorth\nodeName)-(0.85,0)$) (l\nodeName) {\footnotesize$l=\l$};
                        \node[draw=none,rounded corners,minimum width=4.1em] at ($(phSouth\nodeName)-(0.85,0)$) (tau\nodeName) {\footnotesize$\jaroTau^{\l}=\t$};
                        \node[nodecount] (desc\nodeName) at ($(\nodeName.north west) + (0.1em, -0.1em)$) {\textbf{\count}};
                    }
            \end{tikzpicture}
        }
    }
    \caption{PUMA trie for $\{$Chimpanzee, Chimp, Chinchilla, Cobra, Puma$\}$, $\winklerTau = 0.9$; single-entry nodes are skipped (black).}
    \label{fig:trie}
    \Description{Illustration of a PUMA trie for the string set {Chimpanzee, Chimp, Chinchilla, Cobra, Puma} at a Jaro-Winkler threshold of 0.9. The trie has five depth levels (from l=0 to 4), with each level annotated by its depth and adaptive Jaro threshold tau_J, decreasing from 0.9 to 0.833. Nodes are represented as rectangles containing grouped strings, where distinct colors distinguish different subtrees to indicate evaluation between different colored groups at this level. Single-entry nodes appear as solid black rectangles, and directed edges are labeled with the advancing character.}
\end{figure}

\begin{example}

    \newcommand{\inlinecirclenode}[1]{
        \!\tikz[baseline]{%
            \node[draw,circle,inner sep=0.75pt,anchor=base] {#1};
        }\!
    }
    
    Figure~\ref{fig:trie} shows the PUMA trie for $S = \{s_1$:~Cobra, \allowbreak $s_2$:~Chinchilla, \allowbreak $s_3$:~Chimpanzee, \allowbreak $s_4$:~Chimp, \allowbreak $s_5$:~Puma$\}$.   
    For nodes at level $i$, we use the size filter with the depth-specific Jaro threshold $\jaroTau^i$ to determine the pre-candidates.
    In post-order traversal, node \inlinecirclenode{1} represents $i=4$ with prefix \enquote{CHIM} and identifies $T_1 = \{(s_3, s_4)\}$ as partial pre-candidates for $\jaroTau^4$ (self-join on the white node entries; no pushdown possible at $i = l_{\max}$).
    At node \inlinecirclenode{2}, $i=3$, prefix \enquote{CHI}, this pair is skipped as it has already been processed at node \inlinecirclenode{1} (\enquote{M} is shared at pos.\ $i + 1 = 4$).
    A join between red and green node entries yields pairs with common prefix length $l=3$, which are evaluated under $\jaroTau^3$.
    Pair $(s_3, s_2)$ is the only one of those pairs satisfying the size filter with $\jaroTau^3$, and thus $T_3 = \{(s_3, s_2)\}$.
    This processing logic continues upwards.
    Node \inlinecirclenode{3} skips all pairs ($l = 2$); nodes \inlinecirclenode{4} and \inlinecirclenode{5} form only pairs with $s_1$ ($l = 1$) and $s_5$ ($l = 0$), respectively.
    Hence, $T_3 = \emptyset$, $T_4 = \{(s_4, s_1)\}$, and $T_5 = \{(s_1, s_5), (s_4, s_5)\}$.
    The overall set of pre-candidates is $T = \bigcup T_i = \{(s_3, s_4), (s_3, s_2), (s_4, s_1), (s_1, s_5), (s_4, s_5)\}$.
\end{example}

\section{Fast and Robust Overlap Filtering} \label{sec:overlap-join} %

For Jaro-Winkler similarity joins, Dressler et al.~\cite{Dressler:OnEfficientExecutionBoundedJaroWinklerDistances} observe that the overlap filter (cf.\ Lemma~\ref{lem:overlap-filter}) significantly reduces the number of Jaro-Winkler computations.
While the state of the art computes the overlap of two ordered sets using a merge-based approach~\cite{Mann:EmpiricalEvaluationSetSimilarityJoin}, there exist techniques for specific scenarios. %
Small universes favor \textit{exact bitmaps,} where a bit position encodes a token in $\setuniverse$.
The overlap of two exact bitmaps can be computed in time $\mathcal{O}(|\setuniverse|)$ using bitwise $\mathtt{AND}$ (denoted $\land$) followed by a popcount instruction (denoted $\mathtt{POPCNT}$), which counts the number of set bits in the resulting bitmap.
In practice, $|\setuniverse|$ often exceeds the standard 64-bit CPU register size.
Even when using Single Instruction, Multiple Data (SIMD) instructions that operate on wider registers of 128 (SSE), 256 (AVX/AVX2), or 512 bits (AVX-512), the universe size may still be too large.
If the bitmap size exceeds the register size, multiple instructions are required to compute the overlap, leading to increased runtime.
To this end, Sandes et al.~\cite{Sandes:BitmapFilterSpeedingUpExactSetSimilarityJoinsBitwiseOperations} propose \textit{approximate bitmaps} that hash tokens into fixed-size bitmaps.
They employ several collision policies and derive an upper bound on the overlap based on the number of distinct bits:
\begin{align} \label{eq:bitmap-upper-bound-Xor}
    |r \cap s| \leq \left \lfloor \frac{|r| + |s| - \pc{B_r \oplus B_s}}{2} \right \rfloor
\end{align}
where $\oplus$ denotes bitwise $\mathtt{XOR}$ and $B_r$, $B_s$ refer to the approximate bitmaps of sets $r$, $s$, respectively.
A pair $(r,s)$ is pruned if the upper bound is below the overlap threshold.
However, we identify major limitations of approximate bitmaps:
(1) Tokens are assumed to be uniformly distributed, whereas real-world string datasets are often highly skewed~\cite{Cantone:FrequencyCharactersNaturalLanguageTexts}.
(2) Effectiveness and efficiency heavily depend on the similarity threshold and dataset characteristics like the universe size and the set size distribution.
While previous work~\cite{Bellas:EmpiricalEvaluationExactSetSimilarityJoinTechniquesGPUs} has identified \enquote{sweet spots}, performance degrades if the number of tokens increases, as false positives are more likely.
In addition, enlarging the bitmaps implies a large memory footprint, excessive memory accesses, and more cache misses.

We present a novel bitmap blueprint that combines an exact encoding for frequent tokens with a hash encoding for rare tokens.
This layout provides both a lower and an upper bound on the overlap to prune the search space effectively.
We also study the optimal bitmap layout for robust performance across various scenarios.

\subsection{Dual-Bound Intersection Bitmaps}

We introduce the \emph{dual-bound intersection bitmap (\dib{})}, which combines desirable characteristics of exact and approximate bitmaps in a single bitmap blueprint.
A \dib{} is a fixed-size bitmap that is divided into $n$ exact and $k$ hashed bits.
This provides several benefits for threshold-based overlap checks:
(1) The \emph{exact part} yields a lower bound on the overlap.
(2) In combination, both parts yield an upper bound on the overlap.
(3) \dib{}s have a \emph{fixed size} that can be chosen to align with CPU register size.
In the following, we assume two sets $r$, $s$ with the corresponding \dib{}s $B_r$, $B_s$.

\begin{wrapfigure}{r}{0.5\linewidth}
    \contourlength{2.4pt}
    
    \centering
    \begin{tikzpicture}[semithick]
        \node[draw, minimum width=0.15\textwidth, minimum height=1.5em, pattern=bricks, pattern color=blue!75!black] at (0,0) (dib-exact) {\contour{white}{\textbf{$\boldsymbol{n}$ exact bits}}};
        \node[right=-0.025 of dib-exact, draw, minimum width=0.25\textwidth, minimum height=1.5em, pattern=north east lines, pattern color=red!80!black] (dib-approx) {\contour{white}{\textbf{$\boldsymbol{k}$ hashed bits}}};
        \node[left=0 of dib-exact.west] (dib-name) {$\boldsymbol{B_r:}$};
        
        \draw[decoration={brace,mirror,raise=2pt},decorate]
        (dib-exact.south west) -- node[below=4pt] {$\boldsymbol{\exactsimple{r}}$} (dib-exact.south east);
        \draw[decoration={brace,mirror,raise=2pt},decorate]
        (dib-approx.south west) -- node[below=4pt] {$\boldsymbol{\heuristicsimple{r}}$} (dib-approx.south east);
        
        \draw[decoration={brace,mirror,raise=16pt},decorate]
        (dib-exact.south west) -- node[below=18pt] {\textbf{fixed size $\boldsymbol{(n+k)}$}} (dib-approx.south east);
        
        \draw[<-,stealth-,dotted,gray!75!black,text=gray!75!black] (dib-exact.north east) to ++(0,0.25) to ++(0.5,0) node[right] {\textbf{shifting boundary}};
        \draw[line width=2mm] (dib-exact.north east) to (dib-exact.south east);
    \end{tikzpicture}
    \caption{Dual-bound intersection bitmap concept.}
    \label{fig:dib}
    \Description{The figure shows a bitmap labeled B sub r, split into two parts: n exact bits on the left, shown with a brick pattern, and k hashed bits on the right, shown with diagonal lines. Braces indicate the size of each part and the fixed total size of n plus k bits. An arrow points to the boundary between the two parts, showing that it can shift to change the allocation of bits while keeping the overall bitmap size unchanged.}
\end{wrapfigure}
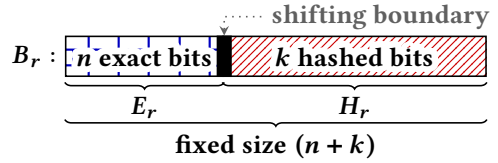

\paragraph{\textbf{Construction}}
The total bitmap size $(n + k)$ and the number of directly encoded bits $n$ are tunable parameters, and Section~\ref{sec:optimal-bitmap-sizes} provides a cost model to derive them automatically.
For a given $n$, we partition the tokens in $\setuniverse$ into two disjoint subsets: $\setuniverse_E$ contains the $n$ most frequent tokens, and $\setuniverse_H$ contains all remaining tokens.
For a set $r$, the first $n$ bits of $B_r$ form the \emph{exact part} $\exactsimple{r}$ and directly encode the frequent tokens in $r \cap \setuniverse_E$ using a one-to-one mapping.

Tokens in $r \cap \setuniverse_H$ are encoded in the \emph{hashed part} $\heuristicsimple{r}$, a space-efficient encoding hashing these tokens into the remaining $k$ bits of $B_r$.
Since $k$ is typically smaller than $|\setuniverse_H|$, collisions are unavoidable.
In case of a collision, several \emph{collision policies} can be applied to resolve it~\cite{Sandes:BitmapFilterSpeedingUpExactSetSimilarityJoinsBitwiseOperations}:
(1) \emph{Linear Probing (\collin):} If a token hashes to an already occupied bit, the following bits are probed sequentially until an empty bit is found and set.
(2) \emph{\colxor:} If a token hashes to an already occupied bit, the bit is flipped using $\mathtt{XOR}$.
(3) \emph{\colset:} A token that hashes to a bit sets the bit regardless of its previous value.

Additionally, we construct a \emph{fallback bitmap} $\bitmaprecsimple{r}$ of size $(|\setuniverse| - n)$, which encodes all tokens in $r \cap \setuniverse_H$ using a one-to-one mapping. 
This provides an exact representation of the hashed part and allows us to compute the precise overlap if both lower and upper bound are inconclusive.

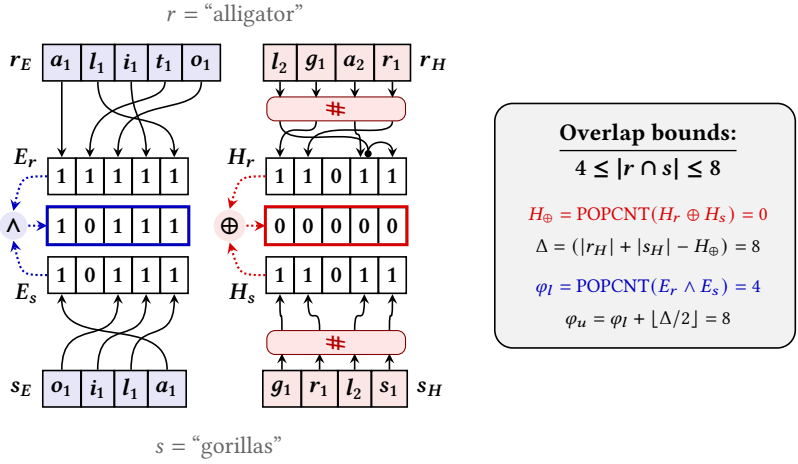
\begin{figure}[t]
    \centering
    {\small
\begin{tikzpicture}[semithick]
    \tikzset{field/.style={draw,minimum height=1.525em}}
    \tikzset{bit/.style={field,minimum width=1.1em}}
    \tikzset{token/.style={field,minimum width=1.35em,fill opacity=0.1,text opacity=1}}
    \tikzset{exact/.style={token,fill=blue!75!black}}
    \tikzset{hashed/.style={token,fill=red!85!black}}
    \tikzset{hashfunc/.style={
        rounded corners=0.35em,
        minimum width=5.8em,
        minimum height=1em,
        fill=red!10,             %
        draw=red!50!black,       %
        thin,                    %
        text=red!70!black,       %
        inner sep=0.075em
    }}
    \tikzset{ptr/.style={-stealth,semithick}}
    \tikzset{flowptr/.style={-stealth,thick}}
    \tikzset{opptr/.style={ptr,thick,densely dotted}}
    \tikzset{opbasic/.style={draw=none,inner sep=0.1em}}
    \tikzset{op/.style={opbasic,circle,fill opacity=0.1,text opacity=1}}
    \tikzset{opAND/.style={op,fill=blue!75!black}}
    \tikzset{opXOR/.style={op,fill=red!85!black}}

    \node[exact] at (0,0) (r1) {$\boldsymbol{a_1}$};
    \foreach \ti/\t [count=\xi from 1] in {2/l_1, 3/i_1, 4/t_1, 5/o_1}{
        \node[exact,right=-0.015 of r\xi] (r\ti) {$\boldsymbol{\t}$};
    }
    \node[draw=none,left=0 of r1] (rHdesc) {$\boldsymbol{r_E}$};
    \node[hashed,right=0.55 of r5] (r6) {$\boldsymbol{l_2}$};
    \foreach \ti/\t [count=\xi from 6] in {7/g_1, 8/a_2, 9/r_1}{
        \node[hashed,right=-0.015 of r\xi] (r\ti) {$\boldsymbol{\t}$};
    }
    \node[draw=none,right=0 of r9] (rRightdesc) {$\boldsymbol{r_H}$};

    \node[left=0.25 of r6] (midR) {};
    \node[above=0.25 of midR,text=gray!80!black] (rdesc) {$r = \text{\enquote{alligator}}$};

    \node[bit,below=1 of r1] (Er1) {$\boldsymbol{1}$};
    \foreach \bi/\b [count=\xi from 1] in {2/1, 3/1, 4/1, 5/1}{
        \node[bit,right=-0.015 of Er\xi] (Er\bi) {$\boldsymbol{\b}$};
    }
    \node[above left=-0.25 and 0 of Er1] (Erdesc) {$\boldsymbol{E_r}$};

    \node[bit,below=1 of r9] (Hr5) {$\boldsymbol{1}$};
    \foreach \xi/\bi/\b in {5/4/1, 4/3/0, 3/2/1, 2/1/1}{
        \node[bit,left=-0.015 of Hr\xi] (Hr\bi) {$\boldsymbol{\b}$};
    }
    \node[above left=-0.25 and 0 of Hr1] (Hrdesc) {$\boldsymbol{H_r}$};

    \node[bit,below=0.15 of Er1] (AndE1) {$\boldsymbol{1}$};
    \foreach \bi/\b [count=\xi from 1] in {2/0, 3/1, 4/1, 5/1}{
        \node[bit,right=-0.015 of AndE\xi] (AndE\bi) {$\boldsymbol{\b}$};
    }

    \node[bit,below=0.15 of Hr1] (XorH1) {$\boldsymbol{0}$};
    \foreach \bi/\b [count=\xi from 1] in {2/0, 3/0, 4/0, 5/0}{
        \node[bit,right=-0.015 of XorH\xi] (XorH\bi) {$\boldsymbol{\b}$};
    }

   \draw[blue!75!black, line width=1.1pt, preaction={draw=white, line width=1.pt}] 
    (AndE1.north west) rectangle (AndE5.south east);

    \draw[red!85!black, line width=1.1pt, preaction={draw=white, line width=1.pt}] 
        (XorH1.north west) rectangle (XorH5.south east);

    \node[bit,below=0.15 of AndE1] (Es1) {$\boldsymbol{1}$};
    \foreach \bi/\b [count=\xi from 1] in {2/0, 3/1, 4/1, 5/1}{
        \node[bit,right=-0.015 of Es\xi] (Es\bi) {$\boldsymbol{\b}$};
    }
    \node[below left=-0.25 and 0 of Es1] (Esdesc) {$\boldsymbol{E_s}$};

    \node[bit,below=0.15 of XorH1] (Hs1) {$\boldsymbol{1}$};
    \foreach \bi/\b [count=\xi from 1] in {2/1, 3/0, 4/1, 5/1}{
        \node[bit,right=-0.015 of Hs\xi] (Hs\bi) {$\boldsymbol{\b}$};
    }
    \node[below left=-0.25 and 0 of Hs1] (Hsdesc) {$\boldsymbol{H_s}$};

    \node[exact,below=1 of Es1] (s1) {$\boldsymbol{o_1}$};
    \foreach \ti/\t [count=\xi from 1] in {2/i_1, 3/l_1, 4/a_1}{
        \node[exact,right=-0.015 of s\xi] (s\ti) {$\boldsymbol{\t}$};
    }
    \node[draw=none,left=0 of s1] (sE) {$\boldsymbol{s_E}$};
    \node[draw=none,right=0.215 of s4] (sPh) {\phantom{$\boldsymbol{g_1}$}};
    \node[hashed,right=0.25 of sPh] (s5) {$\boldsymbol{g_1}$};
    \foreach \ti/\t [count=\xi from 5] in {6/r_1, 7/l_2, 8/s_1}{
        \node[hashed,right=-0.015 of s\xi] (s\ti) {$\boldsymbol{\t}$};
    }
    \node[draw=none,right=0 of s8] (sH) {$\boldsymbol{s_H}$};
    \node[below left=0.25 and 0.13 of s6,text=gray!80!black] (sdesc) {$s = \text{\enquote{gorillas}}$};

    \node[hashfunc,above=0.475 of Hr3] (hr) {\Large\rotatebox{90}{\textbf{\#}}};
    \node[hashfunc,below=0.475 of Hs3] (hs) {\Large\rotatebox{90}{\textbf{\#}}};

    \node[opAND,left=0.25 of AndE1] (AndE) {$\boldsymbol{\land}$};
    \node[opbasic,left=0.3 of Er1] (Er1AndE) {};
    \node[opbasic,left=0.3 of Es1] (Es1AndE) {};
    \draw[opptr, blue!75!black] (Er1.west) .. controls (Er1AndE.center) .. (AndE.north);
    \draw[opptr, blue!75!black] (Es1.west) .. controls (Es1AndE.center) .. (AndE.south);
    \draw[opptr, blue!75!black] (AndE.east) to (AndE1.west);

    \node[opXOR,left=0.25 of XorH1] (XorH) {$\boldsymbol{\oplus}$};
    \node[opbasic,left=0.3 of Hr1] (Hr1XorE) {};
    \node[opbasic,left=0.3 of Hs1] (Hs1XorE) {};
    \draw[opptr, red!85!black] (Hr1.west) .. controls (Hr1XorE.center) .. (XorH.north);
    \draw[opptr, red!85!black] (Hs1.west) .. controls (Hs1XorE.center) .. (XorH.south);
    \draw[opptr, red!85!black] (XorH.east) to (XorH1.west);

    \draw[ptr] (r1.south) to ($(r1.south)!0.5!(Er1.north)$) to (Er1.north);
    \draw[ptr] (r2.south) .. controls ($(r2.south)-(0,0.75)$) and ($(Er5.north)+(0,0.75)$) .. (Er5.north);
    \draw[ptr] (r3.south) .. controls ($(r3.south)-(0,0.5)$) and ($(Er4.north)+(0,0.5)$) .. (Er4.north);
    \draw[ptr] (r4.south) .. controls ($(r4.south)-(0,0.375)$) and ($(Er2.north)+(0,0.875)$) .. (Er2.north);
    \draw[ptr] (r5.south) .. controls ($(r5.south)-(0,0.5)$) and ($(Er3.north)+(0,0.875)$) .. (Er3.north);

    \foreach \i in {6, 7, 8, 9}{
        \node[draw=none] at ($(r\i.south)-(0,0.2125)$) (rPh\i) {};
        \node[draw=none] at ($(rPh\i.south)-(0,0.195)$) (HrPh\i) {};
        \draw[ptr] (r\i.south) to (rPh\i.center);
    }
    \draw[ptr] (HrPh7.center) .. controls ($(HrPh7.center)-(0,0.3)$) and ($(Hr1.north)+(0,0.6)$) .. (Hr1.north);
    \draw[ptr] (HrPh9.center) .. controls ($(HrPh9.center)-(0,0.3)$) and ($(Hr2.north)+(0,0.6)$) .. (Hr2.north);
    \draw[ptr] (HrPh8.center) .. controls ($(HrPh8.center)-(0,0.35)$) and ($(Hr4.north)+(-0.05,0.5)$) .. ($(Hr4.north)-(0.05,0)$);
    \draw[ptr,-{Circle[width=0.3em,length=0.3em]}] (HrPh6.center) .. controls ($(HrPh6.center)-(0,0.45)$) and ($(Hr4.north)+(0.05,0.5)$) .. ($(Hr4.north)+(0.05,0)$);
    \draw[ptr] ($(Hr4.north)+(0.05,0)$) .. controls ($(Hr4.north)+(0,0.25)$) and ($(Hr5.north)+(0,0.25)$) .. (Hr5.north);

    \draw[ptr] (s1.north) .. controls ($(s1.north)+(0,0.75)$) and ($(Es3.south)-(0,0.75)$) .. (Es3.south);
    \draw[ptr] (s2.north) .. controls ($(s2.north)+(0,0.75)$) and ($(Es4.south)-(0,0.75)$) .. (Es4.south);
    \draw[ptr] (s3.north) .. controls ($(s3.north)+(0,0.75)$) and ($(Es5.south)-(0,0.75)$) .. (Es5.south);
    \draw[ptr] (s4.north) .. controls ($(s4.north)+(0,0.75)$) and ($(Es1.south)-(0,0.75)$) .. (Es1.south);

    \foreach \i in {5, 6, 7, 8}{
        \node[draw=none] at ($(s\i.north)+(0,0.2125)$) (sPh\i) {};
        \node[draw=none] at ($(sPh\i.north)+(0,0.195)$) (HsPh\i) {};
        \draw[ptr] (s\i.north) to (sPh\i.center);
    }
    \draw[ptr] (HsPh5.center) .. controls ($(HsPh5.center)+(0,0.4)$) and ($(Hs1.south)-(0,0.5)$) .. (Hs1.south);
    \draw[ptr] (HsPh6.center) .. controls ($(HsPh6.center)+(0,0.4)$) and ($(Hs2.south)-(0,0.5)$) .. (Hs2.south);
    \draw[ptr] (HsPh7.center) .. controls ($(HsPh7.center)+(0,0.4)$) and ($(Hs4.south)-(0,0.5)$) .. ($(Hs4.south)-(0,0)$);
    \draw[ptr] (HsPh8.center) .. controls ($(HsPh8.center)+(0,0.4)$) and ($(Hs5.south)-(0,0.5)$) .. ($(Hs5.south)-(0,0)$);

    \node[draw=black, right=1.15 of XorH5, rounded corners=0.75em, inner sep=0.6em, fill=gray!10] (overlapbounds) {
        \renewcommand{\arraystretch}{1.1}
        \begin{tabular}{c}
            \textbf{\underline{Overlap bounds:}} \\
            $\boldsymbol{4 \leq |r \cap s| \leq 8}$ \\[1.2ex]
            \color{red!85!black}\scriptsize$H_\oplus = \mathrm{POPCNT}(H_r \oplus H_s) = 0$ \\
            \scriptsize$\Delta = (|r_H| + |s_H| - H_\oplus) = 8$ \\[1ex]
            \color{blue!75!black}\scriptsize$\varphi_l = \mathrm{POPCNT}(E_r \land E_s) = 4$ \\
            \scriptsize$\varphi_u = \varphi_l + \left\lfloor \Delta/2 \right\rfloor = 8$ \\
        \end{tabular}
    };

\end{tikzpicture}
}
    \caption{\dib{} construction for two strings ${r}$, ${s}$ with lower bound ${\varphi_l}$ and upper bound ${\varphi_u}$; ${n = k = 5}$ (cf.\ Example~\ref{ex:dib-construction-bounds}).}
    \label{fig:heuristic_bitmap_example}
    \Description{The figure illustrates the construction of a dual-bound intersection bitmap (DIB) for two strings, r="alligator" and s="gorillas". The DIB consists of two parts: an exact part and a hashed part. The exact part encodes the most frequent tokens directly, while the hashed part encodes the remaining tokens using a hash function. Furthermore, the figure shows the computation of the lower bound and upper bound of the overlap between the two strings based on their DIBs. The lower bound is derived from the exact parts, while the upper bound is computed using both the exact and hashed parts.} 
\end{figure}

Overall, a \dib{} is defined as a tuple $\langle B_r, \bitmaprecsimple{r} \rangle$, where $B_r$ is the concatenation of the exact and hashed parts, $B_r = \exactsimple{r} \unionbitm \heuristicsimple{r}$. 
Figure~\ref{fig:heuristic_bitmap_example} illustrates the construction for two sets $r$, $s$ ($F_r$ and $F_s$ are omitted).

\begin{lemma}[\dib{} Overlap Bounds]\label{lem:dib-bounds}
    Given two sets $r$, $s$ and the corresponding \dib{}s $B_r = E_r \unionbitm H_r$, $B_s = E_s \unionbitm H_s$, the overlap $|r \cap s|$ is bounded by
    \begin{align}
        \varphi_l \leq |r \cap s| \leq \varphi_l + \left\lfloor (|r_H| + |s_H| - H_\oplus) / 2 \right\rfloor =: \varphi_u
    \end{align}
    where $\varphi_l = \pc{\exactsimple{r} \land \exactsimple{s}}$ and $H_\oplus = \pc{\heuristicsimple{r} \oplus \heuristicsimple{s}}$, while $|r_H|$ and $|s_H|$ denote the number of tokens encoded in $H_r$ and $H_s$, respectively.
\end{lemma}

Since $E_r$, $E_s$ directly encode tokens of $r$, $s$, $\varphi_l$ is a partial overlap and forms a natural lower bound.
$\varphi_u$ is an upper bound as we add an upper bound of the hashed part equivalent to Equation~\ref{eq:bitmap-upper-bound-Xor} to $\varphi_l$.

\begin{example}\label{ex:dib-construction-bounds}
    Figure~\ref{fig:heuristic_bitmap_example} illustrates the \dib{} construction for $r =\text{\enquote{alligator}}$, $s=\text{\enquote{gorillas}}$, $k = 5$, and $n = 5$. %
    The universe $\setuniverse =  \{a_1, a_2, e_1, g_1, i_1, l_1, l_2, n_1,o_1, r_1, s_1, t_1\}$ is split into two disjoint subsets: $\setuniverse_E = \{l_1, o_1, i_1, a_1, t_1\}$ are directly encoded in $\exactsimple{r}$, $\exactsimple{s}$, whereas $\setuniverse_H = \{g_1, r_1, s_1, n_1, l_2, e_1, a_2\}$ are hashed into $\heuristicsimple{r}$, $\heuristicsimple{s}$ with \collin{} as collision policy.
    We derive the lower bound as $\varphi_l = \pc{\exactsimple{r} \land \exactsimple{s}} = 4$, and the upper bound as $\varphi_u = \varphi_l + \left\lfloor (|r_H| + |s_H| - \pc{\heuristicsimple{r} \oplus \heuristicsimple{s}}) / 2 \right\rfloor = 4 + 4 = 8$.
    Note that we overestimate the overlap for the tokens in $\setuniverse_H$, e.g., $a_2$ and $l_2$ collide at bit 4 of $\heuristicsimple{r}$.
    While $l_2$ sets the next free bit in $\heuristicsimple{r}$ (bit 5), $l_2$ hashes to bit 4 in $\heuristicsimple{s}$.
    Hence, $\heuristicsimple{r}$ and $\heuristicsimple{s}$ contain a $1$ at bit 4 due to two different tokens.
    For $4 \leq \theta \leq 8$, our bounds are inconclusive, and we must compute the precise overlap using our fallback, i.e., $|r \cap s| = \varphi_l + \pc{\bitmaprecsimple{r} \land \bitmaprecsimple{s}} = 4 + 3 = 7$ (omitted in Figure~\ref{fig:heuristic_bitmap_example}).
\end{example}

\paragraph{\textbf{Overlap Verification}}

Figure~\ref{fig:heuristic_bitmap_example} exemplifies the bounds computation on a conceptual level.
Contrastingly, Algorithm~\ref{alg:verifyOverlap} shows the optimized implementation of the overlap filter for two sets $r$, $s$ using one bitwise $\mathtt{XOR}$ and a bitmask.
For $r$ (analogously $s$), we assume access to the set size $|r|$ and the number of tokens mapped to the exact $|r_E|$ and the hashed part $|r_H|$, i.e., $|r| = |r_E| + |r_H|$.
First, we compute $D_\oplus = B_r \oplus B_s$, and $\pc{D_\oplus}$ represents the Hamming distance $\Delta$ between $B_r$ and $B_s$ (line~\ref{line:DXorDelta}).
Afterwards, we derive the upper bound $\varphi_u$ by subtracting $\Delta$ from the combined set sizes $|r|+|s|$ and dividing by 2 (line~\ref{line:ub}).
If $\varphi_u$ is below $\theta$, the pair violates the overlap filter and we return false (line~\ref{line:ub_check}).
Otherwise, we mask $D_\oplus$ such that only states of the first $n$ bits are retained in $E_\oplus$ (line~\ref{line:EXor}).
Then, we derive the lower bound $\varphi_l = \lfloor(|r_E| + |s_E| - E_\oplus) / 2\rfloor$, which is logically equivalent to $\pc{E_r \land E_s}$ (line~\ref{line:lb}).
Finally, we return true if the lower bound is satisfied (line~\ref{line:lb_check}); otherwise, we verify the exact overlap using the fallback bitmaps (line~\ref{line:fallback}).

\begin{algorithm}[t]
    \small
    \DontPrintSemicolon
    \caption{Overlap Verification with \dib{}s}
    \label{alg:verifyOverlap}
    \KwIn{$\langle B_r, \bitmaprecsimple{r} \rangle$, $\langle B_s, \bitmaprecsimple{s} \rangle$ with $n$ exact, $k$ hashed bits; threshold $\theta$}
    \KwOut{True if $|r \cap s| \geq \theta$, false otherwise}
    $D_\oplus \gets B_r \oplus B_s$; $\Delta = \pc{D_\oplus}$\;\label{line:DXorDelta}
    $\varphi_u \gets (|r| + |s| - \Delta) \gg  1$\;\label{line:ub}
    \lIf{$\varphi_u < \theta$}{
        \Return \textbf{false}\label{line:ub_check}
    }
    $E_\oplus \gets D_\oplus \land (1^n0^k)$\label{line:EXor} \tcp*[l]{Mask $D_\oplus$ to see only exact bits}
    $\varphi_l \gets (|r_E| + |s_E| - \pc{E_\oplus}) \gg 1$\;\label{line:lb}
    \lIf{$\varphi_l \geq \theta$}{
        \Return \textbf{true}\label{line:lb_check}
    }
    \Return \pc{\bitmaprecsimple{r} \land \bitmaprecsimple{s}} + $\varphi_l \geq \theta$ \label{line:fallback}
    
\end{algorithm}

\subsection{Optimal Bitmap Layout} \label{sec:optimal-bitmap-sizes}

In general, the layout of a \dib{} can be chosen arbitrarily.
A key challenge towards an efficient and effective overlap filter is to find appropriate values for $n$ and $k$, i.e., the number of bits for the exact and the hashed part, under an appropriate collision policy.
We model the expected verification cost and provide an algorithm for deriving the optimal layout given a dataset and similarity threshold. Our hardware-agnostic model requires only statistics collected in a single pass over the dataset.

\paragraph{\textbf{A Cost Model for \dib{}s}}

\newcommand{\basecost}{C_\mathit{base}}
\newcommand{\fbcost}{C_\mathit{fb}}
\newcommand{\ubprob}{P_\mathit{ub}}
\newcommand{\lbprob}{P_\mathit{lb}}
\newcommand{\fbprob}{P_\mathit{f}}

The verification shown in Algorithm~\ref{alg:verifyOverlap} involves three stages with different costs:
We compute (1) an upper bound, conditionally followed by (2) a lower bound and (3) the exact overlap if both bounds are inconclusive.
We refer to these stages as \emph{upper bound check}, \emph{lower bound check}, and \emph{fallback}, respectively.

The costs of a check depend on the  bitmap size (i.e., $n + k$ and $|\setuniverse| - n$) and the underlying architecture, i.e., the number of instructions required to compute the bounds.
These costs are not necessarily linear w.r.t.\ the bitmap size, in particular when SIMD instructions are used.
Thus, we conduct a small offline experiment to measure the \emph{base costs} $\basecost(B)$ for a bitmap $B$ of size $n + k$ (fixed), and the \emph{fallback costs} $\fbcost(|\setuniverse| - n)$ for a fallback bitmap of size $|\setuniverse| - n$.
Using these costs, we model the \emph{expected cost $C(n,k)$} as
\begin{equation} \label{eq:cost-model}
    C(n,k) = \basecost(B) \cdot \bigl(2 - \ubprob(n,k)\bigr) + \fbprob(n,k) \cdot \fbcost(|\setuniverse| - n)
\end{equation}
where $\ubprob(n,k)$ is the probability that a pair is pruned by the upper bound, and $\fbprob(n,k)$ is the probability that both bounds are inconclusive. %
$\ubprob(n,k)$ and $\fbprob(n,k)$ depend on the dataset characteristics, the similarity threshold, and the collision policy.
Our goal is to find $(n^{\ast}, k^{\ast})$ and a collision policy that minimize the expected cost:
\begin{equation}
    (n^{\ast}, k^{\ast}) = \arg\min_{n,k} C(n,k).
\end{equation}

\paragraph{\textbf{Bitmap Overlap Distributions}}

To estimate $\ubprob(n,k)$ and $\fbprob(n,k)$, we model the overlap distribution using random bitmaps.
Assume two random bitmaps $B_r, B_s \in \{0,1\}^m$ with $m$ bits, and let $t_b^r$, $t_b^s$ be the probabilities that bit $b$ is set in $B_r$, $B_s$, respectively.
Since bits are independent, the probability that bit $b$ is set in both $B_r$, $B_s$ is $t_b^r \cdot t_b^s$.
Let $X_b$ be a random variable that is $1$ if bit $b$ is set in both $B_r$ and $B_s$, and $0$ otherwise.
Then, the overlap of $B_r$ and $B_s$ is $X = \sum_{b=1}^m X_b$ and follows a Poisson binomial distribution with $\{t_1^r \cdot t_1^s, \ldots, t_m^r \cdot t_m^s\}$.

\paragraph{\textbf{Probability Estimations}}

Since $\fbprob(n,k) = 1 - \lbprob(n,k) - \ubprob(n,k)$ for a given overlap threshold $\theta$, we need to estimate $\lbprob(n,k) = \prob(L \ge \theta)$ and $\ubprob(n,k) = \prob(U < \theta)$, where $L$ is a random variable for the overlap in the exact part, $H$ is the overlap upper bound in the hashed part, and $U = L+H$ is the \emph{total overlap upper bound.}

Similar to other normalized similarity measures (e.g., Jaccard), the required overlap threshold $\theta$ for Jaro-Winkler varies for each set pair based on their sizes (cf. Lemma~\ref{lem:overlap-filter})~\cite{Xiao:EfficientSimilarityJoinsNearDuplicateDetectionPOSITIONAL,Mann:EmpiricalEvaluationSetSimilarityJoin}.
Moreover, most similarity joins integrate the size filter, which changes the distribution of set pairs reaching overlap verification.
Hence, we optimize for minimal cost of pairs generated by size-filtered algorithms.

We group records of similar size in \emph{size-groups} $\mathcal{S}^1, \dots, \mathcal{S}^m$, where $m$ is user-defined.
A group is defined as $\mathcal{S}^i = \smash{\bigl(l_{\min}^i,\allowbreak\ l_{\mathrm{med}}^i,\allowbreak\ l_{\max}^i,\allowbreak\ \mathcal{P}^i\bigr)}$, where $l_{\min}^i$, $l_{\mathrm{med}}^i$, and $l_{\max}^i$ are the minimum, median, and maximum set sizes in group $i$, and $\mathcal{P}^i$ denotes the associated token probabilities.
Over all non-symmetric group-pairs $(\mathcal{S}^i, \mathcal{S}^j)$, we derive group-pair-specific probabilities $\smash{\ubprob^{ij}(n,k)}$, $\smash{\lbprob^{ij}(n,k)}$, and $\smash{\fbprob^{ij}(n,k)}$ for the set pairs in $\mathcal{S}^i \times \mathcal{S}^j$ that reach the overlap verification:
\begin{enumerate*}
    \item \emph{No set pairs in $\mathcal{S}^i \times \mathcal{S}^j$ satisfy the size filter:} We discard the group pair as it does not contribute to the cost of the overlap verification.
    \item \emph{Set pairs in $\mathcal{S}^i \times \mathcal{S}^j$ might satisfy the size filter:} We compute $\smash{\ubprob^{ij}(n,k)}$, $\smash{\lbprob^{ij}(n,k)}$, and $\smash{\fbprob^{ij}(n,k)}$ based on the token probabilities $\mathcal{P}^i$ and $\mathcal{P}^j$, with a group-specific overlap threshold $\theta_{ij}$ derived from Lemma~\ref{lem:overlap-filter} with median set sizes $l_{\mathrm{med}}^i$ and $l_{\mathrm{med}}^j$.
\end{enumerate*}

For a group pair $(\mathcal{S}^i, \mathcal{S}^j)$, we derive $\smash{\ubprob^{ij}(n,k)} = \mathbb{P}(U^{ij} < \theta_{ij})$ and $\smash{\lbprob^{ij}(n,k)} = \mathbb{P}(L^{ij} \ge \theta_{ij})$, where $L^{ij}, U^{ij}$ are the random variables for the lower bound and the total overlap upper bound for set pairs in $\mathcal{S}^i \times \mathcal{S}^j$.
For independent tokens, both $L^{ij}, U^{ij}$, and the collision policy can be derived from $\mathcal{P}^i, \mathcal{P}^j$.
Let $\setuniverse$ be sorted w.r.t.\ descending global token frequency and $p_j^i$ be the probability that a random set from $\mathcal{S}^i$ contains the $j$-th token.
$L^{ij}$ represents the overlap of the $n$ most frequent tokens and follows a Poisson binomial distribution with $\{p_1^i \cdot p_1^j, \dots, p_n^i \cdot p_n^j\}$, where $p_m^i \cdot p_m^j$ is the probability that the $m$-th token occurs in two independent sets from $\mathcal{S}^i$ and $\mathcal{S}^j$.

In case of $\smash{H^{ij}}$, estimating the distribution is more complex. %
Recall that $t_b^i$ is the probability that bit $b$ is set in the hashed part of a random set from $\mathcal{S}^i$.
We define $\smash{D^{ij}}$ as the random variable for the Hamming distance between the hashed parts of two random sets from $\smash{\mathcal{S}^i \times \mathcal{S}^j}$.
Since the probability that \emph{exactly one} hashed bitmap contains a $1$ at a given bit $b$ is $\smash{x_b^{ij} = t_b^i \cdot (1 - t_b^j) + (1 - t_b^i) \cdot t_b^j}$, $\smash{D^{ij}}$ follows a Poisson binomial distribution with $\{x_1^{ij}, \dots, x_k^{ij}\}$.
Moreover, we model the number of tokens encoded in the hashed part of two random sets from $\smash{\mathcal{S}^i \times \mathcal{S}^j}$ as independent random variables $T^i$ and $T^j$ under a Poisson binomial distribution with $\{p_{n+1}^i, \dots, p_{|\setuniverse|}^i\}$ and $\{p_{n+1}^j, \dots, p_{|\setuniverse|}^j\}$, respectively.
Based on this, $\smash{H^{ij} = \lfloor \tfrac{T^i + T^j - D^{ij}}{2}\rfloor}$.

The probability that bit $b$ is set in the hashed part of a random set in $\mathcal{S}^i$ depends on the collision policy.
Let $I = \{n+1, \dots, |\setuniverse|\}$ be the indexes for tokens in the hashed part.
Then, $\smash{t_b^i = \frac{1}{2} - \frac{1}{2}\prod_{l\in I, h(l) = b } (1 - 2p_{l}^i)}$ for \colxor-based bitmaps, $\smash{t_b^i = 1 - \prod_{l\in I, h(l) = b } (1 - p_{l}^i)}$ for \colset-based bitmaps.
In contrast, $t_b^i$ for linear probing depends on 
\begin{enumerate*}
    \item the probabilities of the tokens that hash to bit $b$, and 
    \item the probability that bit $(b-1 \bmod k)$ is set.
\end{enumerate*}
To the best of our knowledge, no closed-form solution exists for these circular dependencies; hence, we use Monte-Carlo simulation to estimate the probabilities.

Finally, we derive $U^{\smash{ij}} = L^{\smash{ij}} + H^{\smash{ij}}$, $\ubprob^{\smash{ij}}(n,k) = \prob(U^{\smash{ij}} < \theta_{ij})$, $\lbprob^{\smash{ij}}(n,k) = \prob(L^{\smash{ij}} \geq \theta_{ij})$, and $\fbprob^{\smash{ij}}(n,k) = 1 - \lbprob^{\smash{ij}}(n,k) - \ubprob^{\smash{ij}}(n,k)$ for a given group pair $(\mathcal{S}^i, \mathcal{S}^j)$.
Furthermore, we compute the conditional probabilities for pairs that reach the overlap verification by normalizing the probabilities w.r.t.\ the non-discarded group pairs.
Let $q_i$ be the probability of a randomly selected set to belong to $\mathcal{S}^i$.
Then, $q_{ij}$ with $q_{ij}=2 q_i q_j$ for $i \neq j$ and $q_{ii} = q_i^2$ is the probability of a set pair being in $\mathcal{S}^i \times \mathcal{S}^j$.
For the set of indexes for non-discarded group pairs, $V$, and $\mathcal{Q} = \sum_{(i,j)\in V} q_{ij}$, the overall probabilities are:
\begin{equation}
    \fbprob(n, k) = \frac{1}{\mathcal{Q}} \sum_{(i,j) \in V} \fbprob^{ij}(n, k) \cdot q_{ij}
\end{equation}
After computing the probabilities $\ubprob(n,k)$, $\lbprob(n,k)$ analogously, we compute the expected cost $C(n,k)$ for a given bitmap layout and collision policy based on Equation~\ref{eq:cost-model}.

\begin{figure*}[t]
    \centering
    \includegraphics[width=\linewidth]{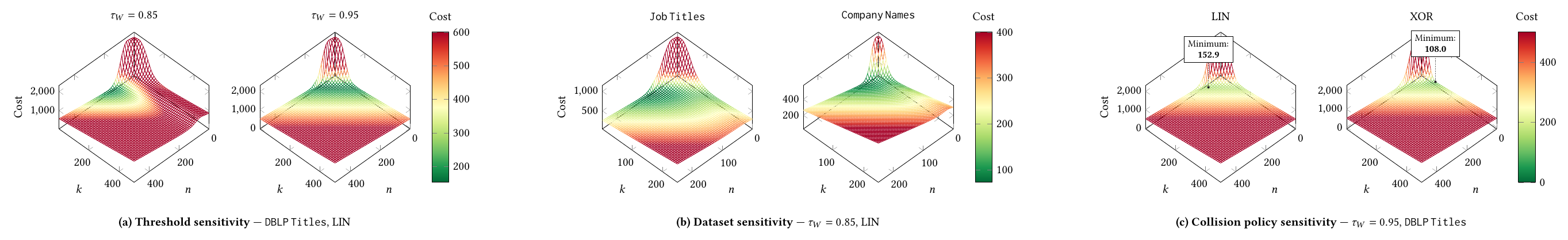}
    \caption{Expected cost ${C(n,k)}$ for different bitmap layouts on different datasets with $\basecost(B) = B$ and $\fbcost(|\setuniverse|) = |\setuniverse|$; Showing sensitivity to (a) the Jaro-Winkler threshold, (b) the token distribution and (c) the collision policy.}
    \label{fig:cost-function-eval}
    \Description{{Six three-dimensional surface plots show the expected cost C(n,k) for varying bitmaps sizes n and k. Each plot uses a color scale, from green for lower costs to red for higher costs. Panel (a) compares Jaro-Winkler thresholds of 0.85 and 0.95 on the DBLP Titles dataset using LIN. Panel (b) compares the Job Titles and Company Names datasets at a threshold of 0.85 using LIN, illustrating how different token distributions affect the cost landscape.
    Panel (c) compares the LIN and XOR collision policies on DBLP Titles at a threshold of 0.95. Both surfaces show how the cost changes with n and k, with marked minimum costs of 152.9 for LIN and 108.0 for XOR, respectively. Overall, the plots show how the expected cost varies with bitmap layout, threshold, token distribution, and collision policy.}}
\end{figure*}

\phantomsection
\label{par:parameter-search}
\paragraph{\textbf{Parameter Search}}

To find the optimal bitmap layout, we perform a grid search over bitmap sizes $B\in[B_{min}; B_{max}]$.
We partition the dataset into contiguous, equiprobable size-groups with roughly the same number of sets.
Then, $C(n,k)$ is evaluated for pairs $(n,k)$ that satisfy $n + k = B$ with $n \in \{i\eta \mid i \in \mathbb{N}_0, i\eta < B\}$ and a user-defined step size $\eta$.
We keep track of the parameters $(n^{\ast}, k^{\ast})$, the associated collision policy, and the expected cost $C_{min}$ for the optimal layout.

We leverage an \emph{early termination criterion} to accelerate the search procedure. Assume that we have just evaluated all bitmaps of size ${<}B$; $(n^{\ast}, k^{\ast})$ with $n^{\ast} + k^{\ast} < B$ stores the current best layout. 
We want to decide whether bitmap layouts $(n,k)$ of size $n+k\geq B$ might have costs $C(n, k)<C(n^{\ast}, k^{\ast})$. In the best case, $\fbprob(n, k) = 0$ and $\ubprob(n, k) = 1$. Hence, $\basecost(B)\leq C(n, k)$. Thus, we can terminate if $\basecost(B) \geq C(n^{\ast}, k^{\ast})$. In practice, we only evaluate bitmap sizes that are multiples of the register size (commonly multiples of $64$).

\begin{example}
    Figure~\ref{fig:cost-function-eval} shows the expected cost $C(n, k)$ for different datasets, thresholds, and collision policies. In general, the expected cost first drops sharply with increasing $n+k$, corresponding to a drop in fallback probability $\fbprob$. Eventually, increasing $n+k$ only increases the cost $\basecost(n+k)$ without meaningfully impacting $\ubprob$ and $\fbprob$. For \texttt{DBLP Titles} on $\winklerTau{=}0.85$, the lower bound is particularly effective, favoring higher values of $n$ compared to $k$.
\end{example}

\paragraph{\textbf{Generalization}}

Our cost model assumes the exact overlap verification to be computed using bitmaps of size $(|\setuniverse| - n)$.
Depending on the dataset characteristics, other verification techniques may be chosen, e.g, small sets with a large universe may favor a merge-based intersection~\cite{Mann:EmpiricalEvaluationSetSimilarityJoin} over sparse bitmaps.
In this regard, our cost model can be conveniently adapted by replacing $\fbcost(|\setuniverse| - n)$ with the cost of the alternative verification technique.
Finally, our cost model can be easily adapted to other similarity measures by computing the overlap thresholds and evaluating the size filter w.r.t\ the alternative similarity measure.

\section{Practical Linear-Time Jaro-Winkler Similarity} \label{sec:jw-join}

This section presents \linearver{} (\textbf{J}aro \textbf{U}sing \textbf{M}erge \textbf{P}ass), \relwork{an efficient, merge-based algorithm} for the Jaro-Winkler similarity between two strings, based on a declarative definition of Jaro mappings.

\subsection{Declarative Jaro Mapping} \label{sec:decl-jaro-mapping}

In this section, we briefly revisit the Jaro similarity and introduce the first declarative definition of a Jaro mapping, which helps us to develop an efficient Jaro-Winkler similarity join framework.

\begin{definition}[Valid Mapping]
    \label{def:valid-mapping}
    A mapping $M\subseteq [|r|] \times [|s|]$ between the characters of two strings $r$, $s$ is \emph{valid} for a given window $\omega\geq 0$ if $M$ is a partial one-to-one mapping, i.e., for all pairs $(a,x), (b,y)\in M: a = b \Leftrightarrow x = y$, and the following properties hold for all $(a,x)\in M$:
    \begin{enumerate}
        \item $r[a]=s[x]$ (character identity);
        \item $|a-x|\leq \omega$ (window constraint)
    \end{enumerate}
\end{definition}

\begin{definition}[Jaro Mapping] \label{def:jaro-mapping}
    A mapping $M$ between the characters of two strings $r$, $s$, $L=\max\{|r|,|s|\}$, is a \emph{Jaro mapping} if 
    \begin{enumerate}
        \item $M$ is valid for window $\max\{\left\lfloor L/2 \right\rfloor - 1, 0\}$
        \item the cardinality $|M|$ of $M$ is maximal;
        \item among all valid mappings with maximal cardinality, $M$ minimizes the cost $m=\sum_{(i,j)\in M}(i+j)$.
    \end{enumerate}
\end{definition}

For a Jaro mapping $M$ of two strings $r$, $s$, the \emph{matched substrings} $\hat{r} = \hat{s}$, are the substrings of $r$, $s$ that contain only mapped characters: $r[a] \in \hat{r} \Leftrightarrow \exists x: (a, x) \in M$ and $s[x] \in \hat{s} \Leftrightarrow \exists a: (a, x) \in M$.

\begin{lemma}[Jaro Mapping Uniqueness] \label{lem:unique-jaro-mapping}
    For any two strings $r$, $s$, all Jaro mappings of $r$, $s$ result in identical matched substrings $\hat{r}$, $\hat{s}$.
\end{lemma}

\begin{corollary} \label{cor:unique-Jaro-similarity}
    The Jaro similarity $\jaro(r, s)$ %
    is unique.
\end{corollary}

\begin{theorem} \label{thm:original-greedy-jaro-mapping}
    The original technique~\cite{jaro-1978-original_Jaro_similarity} computes a Jaro mapping.
\end{theorem}

\subsection{\linearver: Jaro Using Merge Pass} \label{sec:jaro-winkler-similarity-computation}

\relwork{Most} approaches for computing the Jaro-Winkler similarity between two strings $r$, $s$ are based on Winkler's original source code~\cite{Winkler:JWComputation}.
Computing the Winkler boost requires determining the common prefix of two strings, which can be done in $\mathcal{O}(l_{max})$ time.
The bottleneck is the computation of the Jaro similarity, i.e., counting the matches and transpositions.
While highly optimized implementations of the quadratic algorithm using bitwise operations~\cite{RapidFuzz:2025} exist,
all algorithms based on the original greedy algorithm~\cite{jaro-1978-original_Jaro_similarity}
suffer from a fundamental problem: matches are identified by comparing the characters of $r$ to a window of $w = \lfloor \max\{|r|, |s|\} / 2 \rfloor - 1$ characters in $s$.
For $|r| \le |s|$, this incurs a quadratic number of comparisons, as the number of character comparisons satisfies $|r| \cdot w \le \sum_{i=1}^{|r|}\sum_{j=\max\{1,i-w\}}^{\min\{|s|,i+w\}} 1 \le |r|\cdot |s| = \mathcal{O}(|r|\cdot|s|)$, where $|r| \cdot w = \Omega(|r|\cdot|s|)$.
\relwork{We leverage our Jaro Mapping from Definition~\ref{def:jaro-mapping} and propose \linearver{}, a merge-based approach to compute Jaro-Winkler in linear time for bounded alphabets \alphabet.}%

\paragraph{\textbf{Outline}}

To identify all matches, we sort the characters in both strings by value ($val$) and original character position ($pos$; to resolve ties).%
\footnote{\relwork{Depending on the alphabet size $|\alphabet|$ and the string length $n$, we use either counting sort or comparison-based sorting. We refer to the complexity analysis below for details.}}
As a result, all identical character instances are grouped together, and we can efficiently identify matches by processing both sorted strings in a merge-like fashion.
We encode original character positions to test whether matches are within window $w$ and to compute the number of transpositions.
Algorithm~\ref{alg:verifyJW} outlines the verification, consisting of an initialization, a \emph{Matching} phase, and a \emph{Transposition} phase.

{\begin{algorithm}[h]
    \small
    \DontPrintSemicolon
    \caption{Linear-Time Jaro Similarity}
    \label{alg:verifyJW}
    \KwIn{Two sequences \(r', s'\) of pairs \((\mathit{val},\mathit{pos})\) sorted by \(\prec\) %
    }
    \KwOut{The Jaro similarity of \( r, s \) (i.e., the original strings of \(r', s'\)).}
    \( w \gets \lfloor \max\{|r'|, |s'|\}/2 \rfloor - 1\); \( \textit{m} \gets 0 \); \( \textit{t} \gets 0 \); \( I_r \gets 1 \); \( I_s \gets 1 \)\;
    \textbf{initialize} arrays \( \hat{r}[1 \ldots |r'|] \gets 0 \) and \( \hat{s}[1 \ldots |s'|] \gets 0 \)\;
    \While{\( I_r \leq |r'| \) \textbf{and} \( I_s \leq |s'| \)}{\label{alg:verifyJW:matching_start}
    \If{\( r'[I_r].val = s'[I_s].val \)}{ \label{alg:verifyJW:charidentity}
    \If{\( |r'[I_r].\textit{pos} - s'[I_s].\textit{pos}| \le w \)}{\label{alg:verifyJW:matchingcase1-start}
    \( \hat{r}[r'[I_r].\textit{pos}] \gets r'[I_r].val \)\;
    \( \hat{s}[s'[I_s].\textit{pos}] \gets s'[I_s].val \)\;
    \( \textit{m} \gets \textit{m} + 1 \); \( I_r \gets I_r + 1 \); \( I_s \gets I_s + 1 \)\;\label{alg:verifyJW:matchingcase1-end}
    }
    \lElseIf{\( r'[I_r].\textit{pos} < s'[I_s].\textit{pos} \)}{\( I_r \gets I_r + 1 \)}\label{alg:verifyJW:matchingcase2-start}
    \lElse{\( I_s \gets I_s + 1 \)}\label{alg:verifyJW:matchingcase2-end}
    }
    {
    \lIf{\( r'[I_r].val < s'[I_s].val \)}{\( I_r \gets I_r + 1 \)}\label{alg:verifyJW:matchingcase3-start}
    \lElse{\( I_s \gets I_s + 1 \)}\label{alg:verifyJW:matchingcase3-end}
    }
    } \label{alg:verifyJW:matching_end}
    \lIf{\( \textit{m} = 0 \)}{\Return 0}
    
    \( I_r \gets 1 \); \( I_s \gets 1 \); \( m_2 \gets 0 \)\; \label{alg:verifyJW:transposition-init}
    \While{\( I_r \leq |r'| \) \textbf{and} \( I_s \leq |s'| \) \textbf{and} \( m_2 < \textit{m} \)}{\label{alg:verifyJW:transposition_start}
        \lWhile{\( I_r \leq |r'| \) \textbf{and} \( \hat{r}[I_r] = 0 \)}{\( I_r \gets I_r + 1 \)} \label{alg:verifyJW:transpositionskips-start}
        \lWhile{\( I_s \leq |s'| \) \textbf{and} \( \hat{s}[I_s] = 0 \)}{\( I_s \gets I_s + 1 \)} \label{alg:verifyJW:transpositionskips-end}
        \lIf{\( \hat{r}[I_r] \ne \hat{s}[I_s] \)}{\( \textit{t} \gets \textit{t} + 0.5 \)} \label{alg:verifyJW:transposition-neqincrement}
        \( I_r \gets I_r + 1 \); \( I_s \gets I_s + 1 \); \( m_2 \gets m_2 + 1 \)
    } \label{alg:verifyJW:transposition_end}
    \Return{\(\frac{1}{3} \Bigl(\frac{\textit{m}}{|r'|}+\frac{\textit{m}}{|s'|}+\frac{\textit{m}-\textit{t}}{\textit{m}}\Bigr) \)}
\end{algorithm}
}

\paragraph{\textbf{Preprocessing \& Initialization}}

Let $\prec$ be a lexicographical order on characters-position pairs $(val, pos)$, with a strict order on the characters $val \in \alphabet$ and the natural order on the positions $pos \in \mathbb{N}$.
The input strings $r$, $s$ are converted into sequences of character-position pairs $r'$, $s'$ sorted w.r.t.\ to $\prec$.
For joins, we preprocess each string exactly once, e.g., $r' = [(\texttt{a},2), (\texttt{a},4), (\texttt{a},6), (\texttt{b},1), (\texttt{n},3), (\texttt{n},5)]$ for $r = \texttt{banana}$.
Algorithm~\ref{alg:verifyJW} takes $r'$ and $s'$, and initializes two arrays (e.g., using null-entries) to store the characters of $r$ and $s$ that are part of a Jaro match at their original 1-indexed positions.

\paragraph{\textbf{Matching Phase}}

Lines~\ref{alg:verifyJW:matching_start}--\ref{alg:verifyJW:matching_end} find all Jaro matches of $r$ and $s$.
We iterate through $r'$ and $s'$ using two pointers, $1 \leq I_r \leq |r'|$ and $1 \leq I_s \leq |s'|$, and the number of matches $m$ is initialized to $0$.
For $r'[I_r] = (v_r, p_r)$ and $s'[I_s] = (v_s, p_s)$, there are three cases:
(1) If $v_r = v_s$ and the original positions $p_r$, $p_s$ are within the Jaro window $w$, then we record a match, i.e., we increment $m$, $v_r$ is stored in $\hat{r}[p_r]$ and $\hat{s}[p_s]$, and both pointers $I_r$, $I_s$ advance, ensuring each character is matched only once (lines~\ref{alg:verifyJW:matchingcase1-start}--\ref{alg:verifyJW:matchingcase1-end}).
(2) If $v_r = v_s$ but the original positions $p_r$, $p_s$ are not within $w$, then we advance the pointer that points to the smaller position (lines~\ref{alg:verifyJW:matchingcase2-start}--\ref{alg:verifyJW:matchingcase2-end}).
(3) If $v_r \neq v_s$: The pointer that points to the smaller character under $\prec$ is advanced (e.g., $I_r$ advances if $(v_r, p_r) \prec (v_s, p_s)$; lines~\ref{alg:verifyJW:matchingcase3-start}--\ref{alg:verifyJW:matchingcase3-end}) .
We terminate once all elements of either $r'$ or $s'$ have been processed (line~\ref{alg:verifyJW:matching_start}).

\paragraph{\textbf{Transposition Phase}}

Lines~\ref{alg:verifyJW:transposition_start}--\ref{alg:verifyJW:transposition_end} compute the number of transpositions based on $\hat{r}$, $\hat{s}$, and $m$.
Line~\ref{alg:verifyJW:transposition-init} initializes the transposition count $t$ to $0$ and two pointers $I_r$ and $I_s$ to iterate through $\hat{r}$ and $\hat{s}$, respectively.
To find the next match between $\hat{r}$ and $\hat{s}$, we skip unmatched characters (i.e., null-entries) until we encounter a match (lines~\ref{alg:verifyJW:transpositionskips-start}--\ref{alg:verifyJW:transpositionskips-end}).
For a match with $\hat{r}[I_r] \neq \hat{s}[I_s]$, we increment $t$ by $0.5$ (line~\ref{alg:verifyJW:transposition-neqincrement}); otherwise, we advance both $I_r$ and $I_s$ to the next match.
We terminate once all $m$ matches have been processed (line~\ref{alg:verifyJW:transposition_start}).

Then, the Jaro and Jaro-Winkler similarities\footnote{The common prefix length $l$ of $r$ and $s$ can be computed by comparing the first $l_{max} = 4$ characters from $r$ and $s$ in $\mathcal{O}(l_{max})$ time before the Jaro computation.} are computed using Equations~\ref{eq:jaro} and~\ref{eq:jaro-winkler}, respectively. 

\begin{theorem} \label{thm:correctness_lemur}
    Algorithm~\ref{alg:verifyJW} correctly computes the Jaro similarity of two strings $r$ and $s$.
\end{theorem}

\paragraph{\texorpdfstring{\textbf{Positional Filter \relwork{in the Matching Phase}}}{Positional Filter during the Matching Phase}}

In the context of similarity joins, we can further optimize the computation of the Jaro similarity by considering the character positions during the matching phase.
If the current number of matches $m$ and the maximum number of remaining possible matches is below the required number of matches $\theta$ (cf. Lemma~\ref{lem:overlap-filter}), we can safely terminate.
Formally, the maximum number of remaining matches for two sorted sequences $r'$, $s'$ is $\min\{|r'| - i, |s'| - j\}$, where $i$, $j$ are the number of compared characters in $r'$, $s'$, respectively.
Hence, we terminate once $m + \min\{|r'| - i, |s'| - j\} < \theta$.
\relwork{
    While the positional filter works with any lexicographical order $\prec$, sorting the characters by increasing global frequency can lead to earlier termination of the matching phase because rare elements are encountered earlier~\cite{Xiao:EfficientSimilarityJoinsNearDuplicateDetectionPOSITIONAL,Mann:EmpiricalEvaluationSetSimilarityJoin}.
}
\paragraph{\texorpdfstring{\textbf{\relwork{Positional Filter in the Transposition Phase}}}{Positional Filter during the Transposition Phase}}
\relwork{
    We also optimize the transposition phase by terminating early once the Jaro similarity can no longer reach the threshold $\jaroTau$.
    Transpositions only decrease the Jaro similarity and we derive the maximum number of transpositions as
    $t_{max} = \lfloor m (-3 \jaroTau + m (\frac{1}{|r|} + \frac{1}{|s|}) + 1) \rfloor$.
    This still permits a Jaro similarity of at least $\jaroTau$, and we can safely terminate the transposition phase and prune the pair once $t > t_{max}$.
}

\phantomsection
\label{par:complexity-analysis}
\paragraph{\texorpdfstring{\textbf{\relwork{Complexity Analysis}}}{Complexity Analysis}} 
\relwork{
    The time complexity of \linearver\ is dominated by the preprocessing step for Algorithm~\ref{alg:verifyJW}, i.e., sorting the sequences of character-position pairs $r'$ and $s'$.
    Both the matching and the transposition phase require $\mathcal{O}(|r| + |s|)$ time, as each character is processed at most once.
    We analyze the preprocessing step as a function of both the alphabet size $|\Sigma|$ and the string length $n = \max\{|r|, |s|\}$.
    (1) For $|\Sigma| = o(n\log n)$, sorting can be done in $\mathcal{O}(n + |\Sigma|)$ time using counting sort.
    Since the input string is inherently ordered by position and counting sort is stable, a single sort pass w.r.t.\ the character value is sufficient.
    (2) Otherwise, comparison-based sorting requires $\mathcal{O}(n\log n)$ time.
    Hence, the overall time complexity of \linearver\ is $\mathcal{O}(\min{\{n + |\Sigma|, n\log n\}})$.
    With comparison-based sorting, storing the sequences $(r', s', \hat{r}, \hat{s})$ requires $\mathcal{O}(n)$ space.
    When $|\Sigma| = o(n \log n)$, counting sort requires $\mathcal{O}(|\Sigma|)$ additional space for the buckets, resulting in an overall space complexity of $\mathcal{O}(n + |\Sigma|)$. 
    
    In our join setting, each string participates in multiple comparisons and $r'$, $s'$ are precomputed only once, reducing the complexity of subsequent similarity computations to $\mathcal{O}(n)$ time and space.
}

\begin{figure}[t]
    \centering
    {\small
\begin{tikzpicture}[semithick]
    \tikzset{field/.style={
                draw,
                minimum height=1.525em,
                minimum width=1.5em,
                text opacity=1
            }}
    \tikzset{ptr/.style={stealth-stealth,semithick}}
    \tikzset{match/.style={field,fill opacity=0.1,text opacity=1}}
    \tikzset{matchOK/.style={match,fill=green!50!black}}
    \tikzset{matchNOK/.style={match,fill=red!80!black}}
    \tikzset{transposition/.style={circle,inner sep=0.125em,text=white,draw=white,minimum size=1.4em}}
    \tikzset{transpositionOK/.style={transposition,fill=green!50!black}}
    \tikzset{transpositionNOK/.style={transposition,fill=red!80!black}}
    
    \node[field] at (0,0) (rprimepos1) {$\boldsymbol{3}$};
    \foreach \pi/\p [count=\xi from 1] in {2/6, 3/5, 4/2, 5/4, 6/1, 7/7}{
            \node[field,right=-0.015 of rprimepos\xi] (rprimepos\pi) {$\boldsymbol{\p}$};
        }
    \node[draw=none,left=0 of rprimepos1] (rprimeposdesc) {\emph{pos}};
    \node[field,below=-0.015 of rprimepos1] (rprimeval1) {\textbf{A}};
    \foreach \vi/\v [count=\xi from 1] in {2/B, 3/C, 4/D, 5/E, 6/F, 7/G}{
            \node[field,right=-0.015 of rprimeval\xi] (rprimeval\vi) {\textbf{\v}};
        }
    \node[draw=none,left=0 of rprimeval1] (rprimevaldesc) {\emph{val}};
    \node[draw=none,left=0.075 of rprimevaldesc.north west] (rprimedesc) {$\boldsymbol{r'}$};
    
    \node[field,below=0.65 of rprimeval1] (sprimeval1) {\textbf{A}};
    \foreach \vi/\v [count=\xi from 1] in {2/B, 3/C, 4/D, 5/D, 6/F, 7/G}{
            \node[field,right=-0.015 of sprimeval\xi] (sprimeval\vi) {\textbf{\v}};
        }
    \node[draw=none,left=0 of sprimeval1] (sprimevaldesc) {\emph{val}};
    \node[field,below=-0.015 of sprimeval1] (sprimepos1) {$\boldsymbol{5}$};
    \foreach \pi/\p [count=\xi from 1] in {2/3, 3/4, 4/2, 5/7, 6/1, 7/6}{
            \node[field,right=-0.015 of sprimepos\xi] (sprimepos\pi) {$\boldsymbol{\p}$};
        }
    \node[draw=none,left=0 of sprimepos1] (sprimeposdesc) {\emph{pos}};
    \node[draw=none,left=0.075 of sprimeposdesc.north west] (sprimedesc) {$\boldsymbol{s'}$};
    
    \foreach \m in {1, 3, 4, 6, 7}{
            \draw[ptr] (rprimeval\m.south) to node[left=0] {$\m.$} (sprimeval\m.north);
        }
    \node[circle,draw,inner sep=0.085em,above=0.2 of rprimepos1.north east] (matchingphaseno) {$1$};
    \node[draw=none,right=0 of matchingphaseno] (matchingphasedesc) {Matching phase};
    
    \node[matchOK,right=1.5cm of rprimeval7.north east] (rhat1) {\textbf{F}};
    \foreach \vi/\v/\s [count=\xi from 1] in {2/D/matchOK, 3/A/field, 4//field, 5/C/field, 6//field, 7/G/matchOK}{
            \node[\s,right=-0.015 of rhat\xi] (rhat\vi) {\textbf{\v}};
        }
    \node[draw=none,left=0 of rhat1] (rhatdesc) {$\boldsymbol{\hat{r}}$};
    \node[matchOK,right=1.5cm of sprimepos7.north east] (shat1) {\textbf{F}};
    \foreach \vi/\v/\s [count=\xi from 1] in {2/D/matchOK, 3//field, 4/C/field, 5/A/field, 6/G/matchOK, 7//field}{
            \node[\s,right=-0.015 of shat\xi] (shat\vi) {\textbf{\v}};
        }
    \node[draw=none,left=0 of shat1] (shatdesc) {$\boldsymbol{\hat{s}}$};
    \foreach \m/\n in {1/1, 2/2, 7/6}{
            \draw[ptr] (rhat\m.south) to node[transpositionOK] {\scalebox{0.89}{$\boldsymbol{\checkmark}$}} (shat\n.north);
        }
    \foreach \m/\n in {3/4, 5/5}{
            \draw[ptr] (rhat\m.south) to node[transpositionNOK] {$\boldsymbol{\times}$} (shat\n.north);
        }
    
    \node[circle,draw,inner sep=0.085em,above=0.2 of rhat1.north] (transpositionphaseno) {$2$};
    \node[draw=none,right=0 of transpositionphaseno] (transpositionphasedesc) {Transposition phase};
\end{tikzpicture}
}
    \caption{Computing the Jaro-Winkler similarity of strings ${r} = \text{\enquote{FDAECBG}}$ and ${s} = \text{\enquote{FDBCAGD}}$ using \linearver.}
    \label{fig:verification_example}
    \Description{The figure illustrates the two phases of JUMPfor the strings FDAECBG and FDBCAGD. In the matching phase, the characters of each string are sorted alphabetically, with their original positions shown alongside them. Repeated characters are ordered by their original positions. Arrows identify the five matched characters A, C, D, F, and G in both strings. Although B occurs in both strings, it is not matched because its positions are too far apart; E occurs only in the first string. In the transposition phase, the matched characters are shown in their original order in each string. Green check marks indicate characters in the same relative order, F, D, and G, while red crosses indicate characters in different relative order, A and C, leading to one transposition.}

\end{figure}
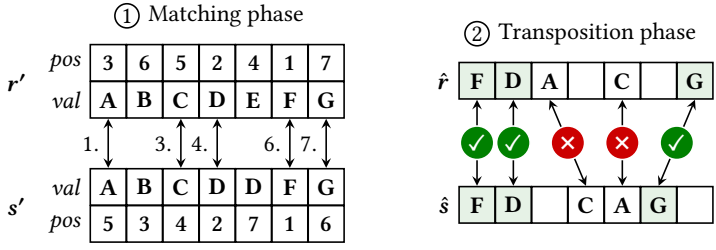

\begin{example}
    \relwork{Figure~\ref{fig:verification_example} illustrates Algorithm~\ref{alg:verifyJW} for two strings $r =$ \enquote{\underline{FD}AECBG}, $s =$ \enquote{\underline{FD}BCAGD} (length 7, $l = 2$ underlined); window size $w = \lfloor \max\{7,7\}/2 \rfloor - 1 = 2$.
        The left-hand side shows the sequences of character-position pairs $r'$ and $s'$ sorted w.r.t.\ $\prec$; the arrays $\hat{r}$ and $\hat{s}$ store matched characters.}
    \emph{(1) Matching Phase.}
    The pointers $I_r$, $I_s$ are initialized to the first elements of $r'$, $s'$, respectively.
    The pair $r'[1]=(\text{A},3)$ and $s'[1]=(\text{A},5)$ is a match as their characters are identical and they are within window $w$.
    Thus, $m=1$, \enquote{A} is stored in $\hat{r}[3]$ and $\hat{s}[5]$, and both pointers advance.
    Next, we compare $r'[2]=(\text{B},6)$ and $s'[2]=(\text{B},3)$. 
    Despite identical characters, this is not a match as the positional distance $|6-3|=3>w$, and we advance $I_s$ ($3<6$).
    In the next pair, we encounter two different characters, $r'[2]=(\text{B},6)$ and $s'[3]=(\text{C},4)$, and advance $I_r$ (\enquote{B} $<$ \enquote{C}).
    We identify further matches $(x,y)\in r'\times s'$: ((C,5), (C,4)), ((D,2), (D,2)), ((F,1), (F,1)), and ((G,7), (G,6)).
    This results in a total number of $m=5$ matches, which are stored in $\hat{r}$, $\hat{s}$ (cf.\ right-hand side of Figure~\ref{fig:verification_example}).
    \emph{(2) Transposition Phase.}
    We determine the number of transpositions by comparing the matched entries of $\hat{r}$ and $\hat{s}$ in a merge-like fashion: \enquote{F}, \enquote{D}, and \enquote{G} occur in positional order, whereas \enquote{A} and \enquote{C} are transposed.
    Hence, two characters occur at different matching positions and the number of transpositions is $t = 2 / 2 = 1.0$.
    \emph{(3) Final Result.}
    We compute Equation~\ref{eq:jaro} for $m = 5$, $t = 1.0$, and $l = 2$:
    $\jaro(r,s) = \frac{1}{3} (\frac{5}{7} + \frac{5}{7} + \frac{5-1.0}{5}) \approx 0.743$.
    Since $\jaro(r,s) \ge 0.7$, we compute Equation~\ref{eq:jaro-winkler} for $p=0.1$: $\jaroWinkler(r,s) = 0.743 + 2 \cdot 0.1 \cdot (1 - 0.743) \approx 0.794$.
\end{example}

\section{\texorpdfstring{\review[23]{Extension to $\bm{\RS}$ Joins}}{Extension to R-S Joins}} \label{sec:rs-joins} %
\review[23]{
    This section describes an extension to the $\RS$ join setting for two different collections $R$, $S$.
    To this end, we discuss (1) how to adapt each algorithmic component, and (2) how a preprocessed relation, $R$, can be reused across multiple joins with different collections.
}
\subsection{\texorpdfstring{\review[3]{Adapting the Components to $\bm{\RS}$ Joins}}{Adapting the Components to R-S Joins}}\label{sec:rs-joins:components}
\review[3]{
    \indent\emph{\textbf{\puma.}} %
    The \puma\ trie is constructed for both collections $R$, $S$ independently.
    Let $n_k$ and $m_k$ denote nodes at depth $k$ in the \puma{} tries over collection $R$ and $S$, respectively.
    Both $n_k$ and $m_k$ refer to the same prefix $p = a_1, \dots, a_k$ with $a_i \in \alphabet$ and $0 \leq k \leq l_{\max}$.
    Furthermore, let $R_k$ and $S_k$ be the set of strings stored in $n_k$ and $m_k$, respectively.
    During the join, we traverse both tries in parallel: for each pair $(n_k, m_k)$, we compute a partial Jaro join $T_k = R_k \simjoin S_k$ with threshold $\jaroTau^k$ (cf.\ Equation~\ref{eq:jaro-threshold}) and verify the pre-candidates w.r.t.\ the Jaro-Winkler similarity.
    If a node of one trie has no corresponding node in the other trie, we skip the partial Jaro join.
    The main difference from the self-join is the pre-candidate generation, which needs to be adapted to the $\RS$ join setting.
    Established techniques such as prefix filtering~\cite{Bayardo:ScalingUpAllPairsSimilaritySearchALLPAIRS} and partition-based filtering~\cite{Deng:EfficientPartitionBasedMethodExactSetSimilarityJoinsPARTALLOC} can be adapted to $\RS$ joins, making them applicable in this context~\cite{Chaudhuri:PrimitiveOperatorSimilarityJoins, Deng:EfficientPartitionBasedMethodExactSetSimilarityJoinsPARTALLOC}.
    To apply size filtering (cf.\ Lemma~\ref{lem:size-filter}), we sort $R$ and $S$ by length independently.
    Deduplication remains unchanged (cf.\ Section~\ref{sec:winkler-trie-duplicates}).
    \paragraph{\textbf{\dib{}}}
    The size groups are computed independently for $R$ and $S$.
    Let $\mathcal{R}^i$ and $\mathcal{S}^j$ be the $i$-th and $j$-th size group of $R$ and $S$, respectively.
    Since $\RS$ joins are asymmetric, we must evaluate all size group pairs $(\mathcal{R}^i, \mathcal{S}^j)$ for which there exists at least one pair $(r, s) \in \mathcal{R}^i \times \mathcal{S}^j$ satisfying the size filter.
    To account for the universes $\setuniverse_R$ and $\setuniverse_S$ of both collections $R$ and $S$, we extend the cost model domain to $\setuniverse = \setuniverse_R \cup \setuniverse_S$.
    All other aspects remain unchanged (cf.\ Section~\ref{sec:optimal-bitmap-sizes}).
    \paragraph{\textbf{\linearver{}}}
    By using the same lexicographical order $\prec$ for both collections $R$ and $S$, \linearver and its complexity remain unchanged (cf.\ Section~\hyperref[par:complexity-analysis]{\ref*{sec:jaro-winkler-similarity-computation}}).
}

\subsection{\texorpdfstring{\review[2]{Preprocessing Reuse Across Joins}}{Preprocessing Reuse Across Joins}}\label{sec:rs-joins:preprocessing-reuse}
\review[2]{
    Significant parts of the preprocessing on $R$ can be reused across different joins and are computed only once:
}
\begin{enumerate*}[label=(\arabic*)]
    \item \review[2]{
              The \puma\ trie on $R$ is built once and reused across all joins.
          }
    \item \review[2]{
              For the cost model, the statistics of $R$ are independent of $\tau_W$ and collected once.
              During joins with a different collection $S$, the statistics of $R$ and $S$ are merged to compute the optimal bitmap layouts.
          }
    \item \review[2]{
              The preprocessing of $R$ for \linearver\ is performed once and reused across all joins.
          }
    \item \review[2]{
              Indexes for pre-candidate generation may also be reusable:
              For size filtering, sorting by length is independent of $\winklerTau$ and reused across different thresholds.
              Prefix-based indexes are built independently of $\winklerTau$ and reused across different thresholds by adjusting the query-time prefix length~\cite{Bayardo:ScalingUpAllPairsSimilaritySearchALLPAIRS}.
              Partition-based indexes remain usable for any $\winklerTau$ larger than a minimum threshold~\cite{Deng:EfficientPartitionBasedMethodExactSetSimilarityJoinsPARTALLOC}.
          }
\end{enumerate*}

\section{Experiments} \label{sec:experiments}\label{sec:exp}

\paragraph{\textbf{Setup}}

We conducted all experiments on an AMD EPYC 9354P CPU with 32 cores and 384 GiB total memory. We implemented all competing algorithms in \texttt{C++} and used \texttt{GCC} 12 with \texttt{CMake} (\texttt{-DCMAKE\_BUILD\_TYPE=Release}) for compilation. For the \rapidfuzz{} algorithm, we use \texttt{rapidfuzz-cpp 3.3.3}~\cite{RapidFuzz:2025}.
Runtime is measured using \texttt{std::chrono}, missing data points correspond to 12-hour timeouts, and we vary $\winklerTau\in \{0.8, 0.83, 0.86, 0.89, 0.92, 0.95\}$.

\paragraph{\textbf{Algorithms}}
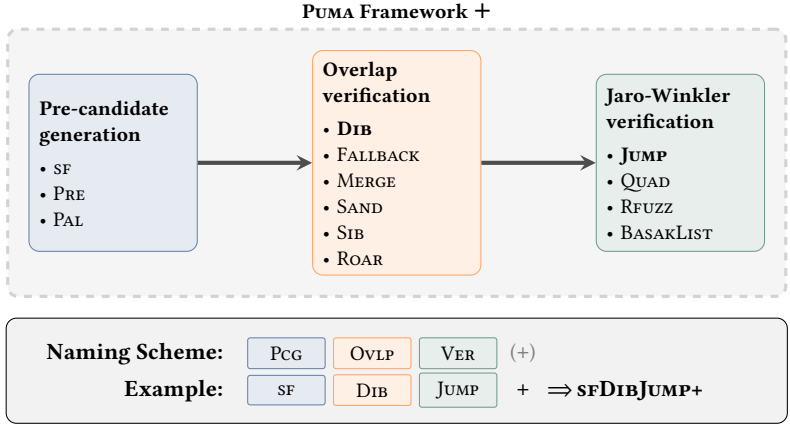
\begin{figure}[t]
    \centering
    \resizebox{0.75\linewidth}{!}{%
        \begin{tikzpicture}[
                font=\footnotesize,
                >=Latex,
                stage/.style={
                        draw,
                        rounded corners,
                        align=left,
                        text width=2.0cm,
                        minimum height=2.4cm,
                        inner sep=4pt
                    },
                pill/.style={
                        draw,
                        rounded corners=1pt,
                        minimum height=4mm,
                        minimum width=1.05cm,
                        align=center,
                        inner xsep=2pt,
                        font=\footnotesize
                    },
                ptr/.style={-stealth, ultra thick, draw=black!70}
            ]

            \newcommand{\stagedistance}{1.55cm}
            
            \node[stage, fill=DBRblue!10, draw=DBRblue!60] (pre) {
                \textbf{Pre-candidate\\generation}\\[1mm]
                \textbullet\ \sizefilter\\
                \textbullet\ \prefix\\
                \textbullet\ \palloc
            };
            
            \node[stage, right=\stagedistance of pre, fill=DBRorange!10, draw=DBRorange!60] (ovlp) {
                \textbf{Overlap\\verification}\\[1mm]
                \textbullet\ \textbf{\dib}\\
                \textbullet\ \fallback\\
                \textbullet\ \merge\\
                \textbullet\ \sandbit\\
                \textbullet\ \sibt\\
                \textbullet\ \roaring
            };
            
            \node[stage, right=\stagedistance of ovlp, fill=DBRgreen!10, draw=DBRgreen!60] (ver) {
                \textbf{Jaro-Winkler\\verification}\\[1mm]
                \textbullet\ \textbf{\linearver}\\
                \textbullet\ \quadver\\
                \textbullet\ \rapidfuzz \\
                \textbullet\ \basakunopt
            };
            
            \draw[ptr] (pre) -- (ovlp);
            \draw[ptr] (ovlp) -- (ver);

            \begin{pgfonlayer}{background}
                \node[
                    draw,
                    rounded corners=4pt,
                    very thick,
                    dashed,
                    fit=(pre) (ovlp) (ver),
                    fill=gray!8,
                    draw=gray!35,
                    inner sep=8pt,
                    label=above:{
                        \textbf{\puma{} Framework\ }{\large{$+$}}
                    }
                ] (puma) {};
            \end{pgfonlayer}

            \node[font=\bfseries, anchor=west] (lbl_scheme) at ([xshift=12pt, yshift=-0.75cm]puma.south west) {\small Naming Scheme:};
            \node[pill, draw=DBRblue!60, fill=DBRblue!10, right=2mm of lbl_scheme] (ns_a) {\textsc{Pcg}};
            \node[pill, draw=DBRorange!60, fill=DBRorange!10, right=1mm of ns_a] (ns_b) {\textsc{Ovlp}};
            \node[pill, draw=DBRgreen!60, fill=DBRgreen!10, right=1mm of ns_b] (ns_c) {\textsc{Ver}};
            \node[text=black!50, right=0.5mm of ns_c] (ns_plus) {($+$)};
            
            \node[font=\bfseries, anchor=east] (lbl_ex) at ([yshift=-0.5cm]lbl_scheme.east) {\small Example:};
            \node[pill, draw=DBRblue!60, fill=DBRblue!10, right=2mm of lbl_ex] (ex_a) {\sizefilter};
            \node[pill, draw=DBRorange!60, fill=DBRorange!10, right=1mm of ex_a] (ex_b) {\dib};
            \node[pill, draw=DBRgreen!60, fill=DBRgreen!10, right=1mm of ex_b] (ex_c) {\linearver};
            \node[right=0.5mm of ex_c] (ex_plus) {\hphantom{(}$+$\hphantom{)}};
            \node[font=\bfseries, right=-0.1cm of ex_plus] (ex_arrow) {$\Rightarrow$};
            \node[font=\bfseries, right=-0.2cm of ex_arrow] (ex_res) {\small \sfdiblinverp{}};

            \coordinate (bottom_left)  at ([yshift=-0.45cm]puma.south west);
            \coordinate (bottom_right) at ([yshift=-0.45cm]puma.south east);

            \begin{pgfonlayer}{background}
                \node[
                    draw,
                    rounded corners=4pt,
                    fill=gray!10,
                    inner xsep=0pt,
                    inner ysep=5pt,
                    fit=(bottom_left) (bottom_right)
                        (lbl_scheme) (lbl_ex) (ex_res)
                ] (bottom_bar) {};
            \end{pgfonlayer}

        \end{tikzpicture}
    }
    \caption{\review[2]{Algorithm naming scheme.}}
    \label{fig:algorithm-overview}
    \Description{Overview of the algorithm naming scheme used in the experiments. The top section presents the three stages of the pipeline: Pre-candidate Generation, Overlap Verification, and Jaro-Winkler Verification, along with the available components for each stage. The bottom section shows how the selected components are combined into a single algorithm name using the corresponding abbreviations (PCG, OVLP, and VER) and a plus sign. An example that shows how these choices form a complete algorithm name, such as sfDibJump+.}
\end{figure}

We compare different algorithmic variants that each implement one or more of the three components of a Jaro-Winkler similarity join (cf.\ Section~\ref{sec:algo-outline}, Figure~\ref{fig:algo-outline}): pre-candidate generation (\textsc{Pcg}), overlap computation (\textsc{Ovlp}), and Jaro-Winkler verification (\textsc{Ver}).
Algorithm labels follow the scheme \textsc{PcgOvlpVer(+)}, where the optional \textsc{+} indicates that the \puma{} framework is used. \review[2]{Figure~\ref{fig:algorithm-overview} summarizes the algorithmic choices and the naming scheme.}

We consider the following \emph{pre-candidate generation} methods:
\begin{enumerate*}
    \item \sizefilter{}: size filtering (cf.\ Lemma~\ref{lem:size-filter}) on length-sorted input strings,
    \item \prefix{}: prefix filtering~\cite{Bayardo:ScalingUpAllPairsSimilaritySearchALLPAIRS}, or
    \item \palloc{}: partition-based filtering~\cite{Deng:EfficientPartitionBasedMethodExactSetSimilarityJoinsPARTALLOC}.
\end{enumerate*}
For \emph{overlap computation} between token sets, we evaluate the following techniques:
\begin{enumerate*}
    \item \dib{} is our proposed solution. Our cost model uses $\eta=4$ and $m=10$ size-groups (no significant benefit for $m>10$ on our datasets and thresholds). We evaluate bitmap sizes in $[64; 512]$ (step size $64$), and use the early termination criterion (cf.\ Section~\hyperref[par:parameter-search]{\ref*{sec:optimal-bitmap-sizes}}) to reduce the search space.\review[3]{The grid search is included in runtime measurements; the one-time dataset and threshold independent offline experiment is not as it is performed only once during compilation.}
    \item \fallback{} uses full exact bitmaps of size $|\setuniverse|$ (cf.\ Section~\ref{sec:overlap-join}).
    \item \merge{} uses sorted arrays of tokens to compute the overlap~\cite{Mann:EmpiricalEvaluationSetSimilarityJoin}.
    \item \sandbit{} by Sandes et al.~\cite{Sandes:BitmapFilterSpeedingUpExactSetSimilarityJoinsBitwiseOperations} (cf.\ Section~\ref{sec:overlap-join}) is a bitmap technique with a mostly data-independent tuning strategy.
    \item \sibt{} by Yang et al.~\cite{Yang:HEROHierarchicalSetPartitioningJoinFrameworkSpeedingUpSetIntersectionOverGraphs} is a partitioning-based overlap algorithm using tree structures.
    \item \roaring{} is a library for compressed bitmaps~\cite{roaringbitmaps}.
\end{enumerate*}

We use one of the following \emph{Jaro-Winkler verification} algorithms:
\begin{enumerate*}
    \item \linearver{} is our linear-time verification (cf.\ Section~\ref{sec:jaro-winkler-similarity-computation}).
    \item \quadver{} is the original quadratic-time verification~\cite{Winkler:JWComputation}.
    \item \rapidfuzz{}~\cite{RapidFuzz:2025} is a \texttt{C++} library that implements the quadratic-time verification using fast bitmaps. \relwork{To the best of our knowledge, we are the first to compare against this highly-optimized baseline algorithm.}
    \item \relwork{\basakunopt{}~\cite{Basak:OnComputingJaroSimilarity,Basak:FastAlgorithmsComputingJaroSimilarity} is a linear-time verification.
              For the per-character lists, we use nested \texttt{std::vector}s, which outperform the proposed 1D-array representation~\cite{Basak:OnComputingJaroSimilarity} in our experiments.}
\end{enumerate*}

For the end-to-end Jaro-Winkler similarity join, we propose \sfdiblinverp{}, \prediblinverp{}, and \paldiblinverp{}. The baseline \sfrfuzz{} leverages size filtering and can be readily implemented using publicly available libraries.
\dressler~\cite{Dressler:OnEfficientExecutionBoundedJaroWinklerDistances} and \keil~\cite{Keil:EfficientBoundedJaroWinklerSimilarityBasedSearch} are specialized trie-based branch-and-bound approaches for Jaro-Winkler joins.
Both use \quadver{} for verification.
For \keil, we add support for incremental trie construction, which substantially improves performance but is not available in the original implementation.
We refer to Section~\ref{sec:related-work} for a detailed description of these algorithms.

\paragraph{\textbf{Datasets}}
\label{sec:exp:setup:datasets}

\begin{wrapfigure}{r}{0.509\linewidth}
    \centering
    \includegraphics[width=\linewidth]{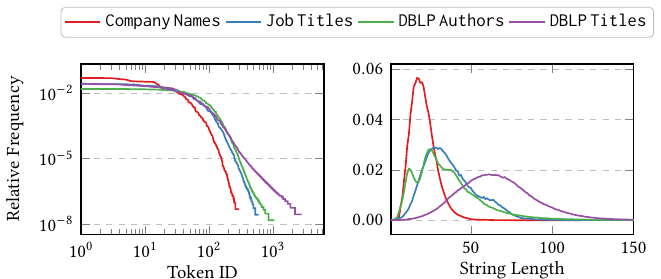}
    \caption{Distribution of string lengths and tokens.}
    \label{fig:dataset_stats}
    \Description{Line plots showing the distribution of string lengths and tokens for four datasets: Company Names, DBLP Titles, DBLP Authors, and Job Titles. The x-axis represents string length or token count, while the y-axis represents the relative frequency. Each dataset is represented by a different colored line, showing how the distributions vary across datasets.}
\end{wrapfigure}

We use four string collections that cover a wide range of relevant characteristics such as string length, universe size, and token frequency distributions.
As Jaro-Winkler is typically applied on short strings, we focus on datasets with an avg.\ string length ${<} 100$ characters.
\texttt{Company\,Names} contains over a million company names published by the UK government~\cite{companieshouse:free-company-data-product}.
\texttt{DBLP\,Titles} and \texttt{DBLP\,Authors} are derived from the DBLP dataset of Tang et al.~\cite{kdd/TangZYLZS08} (DBLP snapshot~\cite{dblp.xml.2023-01-03}) and contain paper titles and author names, respectively.
\texttt{Job\,Titles} aggregates job titles from job portals and applicant tracking systems~\cite{kaggle:7-million-job-titles-companies-locations}.
Table~\ref{tab:datasets} summarizes the basic characteristics; Figure~\ref{fig:dataset_stats} shows the string length and token distributions.

\begin{table}[h]
    \centering
    \caption{Dataset statistics.}
    \label{tab:datasets}
    \small
    \begin{tabular}{c c c c c c}
        \toprule
        Dataset                 & \#Strings & \multicolumn{3}{c}{\#Tokens} & Universe $\setuniverse$                 \\
                                &           & Avg                          & Median                  & 99.9\% &      \\
        \midrule
        \texttt{Company\,Names} & $10^6$    & 20.8                         & 20                      & 57     & 294  \\
        \texttt{DBLP\,Titles}   & $10^6$    & 66.3                         & 64                      & 172    & 1056 \\
        \texttt{DBLP\,Authors}  & $10^6$    & 35.8                         & 31                      & 198    & 2822 \\
        \texttt{Job\,Titles}    & $10^6$    & 37.5                         & 35                      & 115    & 599  \\
        \bottomrule
    \end{tabular}
    \Description{Table showing the statistics of four datasets: Company Names, DBLP Titles, DBLP Authors, and Job Titles. The table includes the number of strings, average number of tokens, median number of tokens, 99.9th percentile of number of tokens, and the size of the universe for each dataset.}
\end{table}

During preprocessing, we remove all non-alphanumeric characters and exact duplicates from the datasets. For approaches that operate on token sets, we use counting tokenization~\cite{Augsten:SimilarityJoinsRelationalDatabaseSystems} (cf.\ Section~\ref{sec:preliminaries}) to transform strings into sets. The universe size (cf.\ Table~\ref{tab:datasets}) is the number of unique tokens in the dataset after tokenization. Due to the overall large runtimes, we use samples of size $10^6$ for all experiments, except for one experiment with particularly high runtimes for competing algorithms. Some experiments (cf.\ Sections~\ref{sec:exp:dib-vs-ovlp} and \ref{sec:exp:jw-verification}) report runtimes for varying string lengths within a single dataset. To this end, we split the dataset into $10$ groups of contiguous string-length ranges, each containing an equal number of strings, and sample $5\cdot 10^4$ random strings from each group.

\subsection{Evaluation of \puma{}}
\label{sec:exp:puma-framework}
We evaluate the impact of the \puma{} framework by comparing \sfdiblinver{}, \paldiblinver{}, and \prediblinver{} with their \puma{} counterparts. 
\begin{figure*}[h]
    \centering
    \includegraphics[width=\figscale\textwidth]{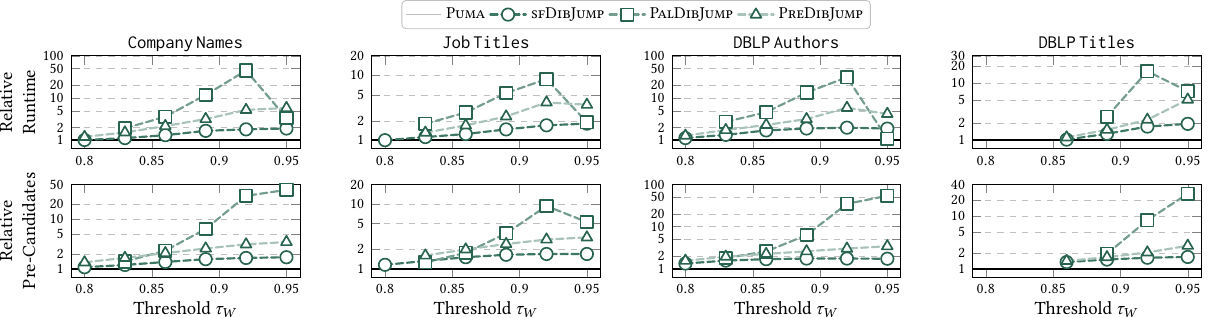}
    \caption{Comparison of different algorithms with and without \puma{} for $10^6$ strings.}
    \label{fig:puma_comp}
    \Description{The line plot shows the runtime and pre-candidate count comparison of different algorithms with and without PUMA for 1 million strings. The x-axis shows the Jaro-Winkler thresholds, while the y-axis shows the normalized runtime and pre-candidate counts relative to PUMA. The algorithms not using PUMA (SF-DIBJUMP, PAL-DIBJUMP, and PRE-DIBJUMP) produce significantly more pre-candidates and are higher runtime than their PUMA counterparts. 
    }
\end{figure*}
Figure~\ref{fig:puma_comp} shows runtimes and pre-candidate counts relative to \puma{}, with the \puma{} variant normalized to 1. On average, the algorithms using \puma{} \sfdiblinverp{}, \paldiblinverp{}, and \prediblinverp{} are faster than their non-\puma{} counterparts by factors of $1.5$, $13.0$, and $2.8$, respectively. This speedup results from a substantial reduction in pre-candidates, by factors of $1.5$, $20$, and $2.9$, respectively. The lower pre-candidate counts result from \puma{} splitting the Jaro-Winkler join into smaller Jaro joins with tighter thresholds $\jaroTau$, yielding tighter filter conditions (cf.\ Section~\ref{sec:winkler-trie}). For \sizefilter{}, the factor-$1.5$ reduction in pre-candidates translates directly into the factor-$1.5$ runtime improvement. For \palloc{} and \prefix{}, the larger pre-candidate reductions translate into slightly smaller runtime improvements, as the overhead of building and probing multiple indexes is amortized by the reduced number of pre-candidates.

\puma{} has the largest impact on \paldiblinver{}.
This is due to a well-known property of \palloc{}, which is highly selective for high similarity thresholds~\cite{Schmitt:ExtensibleRobustSimilarityQueries}.
Since \puma{} increases the join threshold for many string pairs, it substantially boosts the selectivity of \palloc{} for $0.86{<}\winklerTau{<}0.95$.
The only exception is \paldiblinver{} at the highest threshold $\winklerTau{=}0.95$ on \texttt{DBLP\,Authors}, where \puma{} is $20\%$ slower. Although \puma{} still significantly reduces the pre-candidates in this setting,
the overall number of pre-candidates is too low (${<}4\cdot10^7$) to amortize the overhead. The same effect is observed, to a lesser extent, for the other datasets at high thresholds ($\winklerTau{=}0.95$).

Overall, \puma{} substantially improves all pre-candidate generation methods; we therefore focus on the \puma{}-variants henceforth.

\subsection{Evaluation of \dib{}s}
\label{sec:exp:dib-vs-ovlp}

\dib{} uses a data-aware cost model to optimize bitmap size and layout. We first compare \dib{} and \sandbit{} to evaluate their cost models, then assess \dib{}’s scalability against other overlap techniques.

\begin{figure*}[h]
    \centering
    \includegraphics[width=\figscale\textwidth]{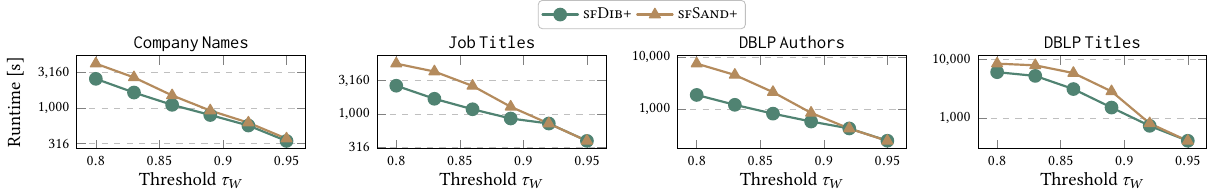}
    \caption{Runtime comparison of the bitmap-based overlap filters \sfdibp{} and \sfsandp{}.}
    \label{fig:dibs_sandes}
    \Description{
        The line plot shows the runtime comparison of the bitmap-based overlap filters SF-DIB+ and SF-SAND+ for 1 million strings. The x-axis shows the Jaro-Winkler thresholds, while the y-axis shows the runtime in seconds. SF-DIBJUMP+ consistently outperforms SF-SANDJUMP+ across all thresholds, with a more pronounced advantage at lower thresholds. The plot illustrates the effectiveness of the data-aware cost model for optimizing bitmap size and layout in SF-DIBJUMP+ compared to the static approach of SF-SANDJUMP+.
    }
\end{figure*}

\paragraph{\textbf{Cost Model Evaluation}}
Figure~\ref{fig:dibs_sandes} shows the performance of \sfdibp{} and \sfsandp{}.
We focus on bitmap performance and apply only the lightweight size filter (\sizefilter{}); no Jaro-Winkler verification is performed.
Both \sfdibp{} and \sfsandp{} dynamically select the hash collision policy (\collin{}, \colxor{}, or \colset{}).
\sfsandp{} selects the bitmap size and collision policy based on the threshold and the avg.\ set size: in our experiments, \sfsandp{} uses \colxor{} for all thresholds $\winklerTau \geq 0.83$ and \colset{} for $\winklerTau = 0.8$ (\collin{} is never used).
\sfsandp{} uses 64-bit bitmaps for datasets with median set size ${<} 50$, and 128-bit bitmaps otherwise.
\sfdibp{} adapts bitmap size, layout, and collision policy to the token distributions based on empirical cost estimations.

Across all datasets and thresholds, \sfdibp{} is on average $71\%$ faster than \sfsandp{}.
For high thresholds ($\winklerTau \geq 0.92$), both \sfdibp{} and \sfsandp{} use similar or identical bitmap layouts, namely \colxor{} with a small or empty exact part for \sfdibp{}, yielding near-identical runtimes.
For lower thresholds ($\winklerTau < 0.92$), \sfsandp{} uses suboptimal bitmap sizes, layouts, and collision policies.
We analyze two aspects in detail:
\begin{enumerate*}
    \item \dib{} provides a lower bound and avoids many overlap computations for truly similar pairs.
          This particularly affects \texttt{DBLP\,Titles} (and to a lesser extent \texttt{Company\,Names}): for lower thresholds, $22\%$ of all generated pre-candidates are handled by the lower bound of \dib{}, avoiding the slower fallback verification.
    \item \dib{} generally selects larger bitmaps and uses \collin{} more often than \sandbit{}.
          This significantly impacts \texttt{Job\,Titles} and \texttt{DBLP\,Authors}: \dib{} uses $128$ bits and $128\text{--}256$ bits with \collin{} on \texttt{Job\,Titles} and \texttt{DBLP\,Authors}, respectively, while \sandbit{} statically selects $64$ bits and $128$ bits with \colset{} and \colxor{}, respectively.
          This results in a significantly lower average fallback probability for \dib{} of $0.3\%$ vs.\ $28\%$ on \texttt{Job\,Titles} and $0.5\%$ vs.\ $18\%$ on \texttt{DBLP\,Authors}.
\end{enumerate*}

\paragraph{\textbf{Scalability}}

\begin{figure*}[h]
    \centering
    \includegraphics[width=\figscale\textwidth]{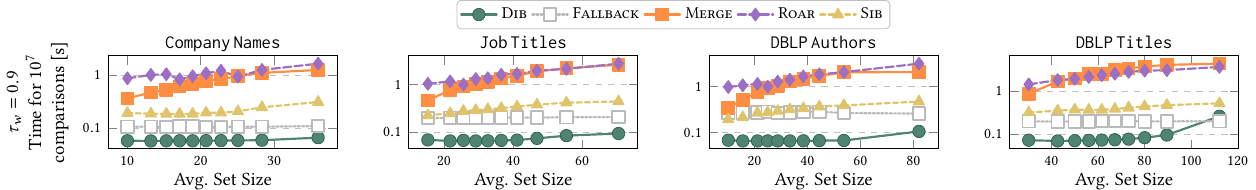}
    \caption{Runtime comparison of different overlap verification algorithms.}
    \label{fig:overlap_verification}
    \Description{
        The line plot shows the runtime comparison of different overlap verification algorithms for 10 million verifications at a Jaro-Winkler threshold of 0.9. The x-axis shows the set size, while the y-axis shows the runtime in seconds per 10 million comparisons. The algorithms compared are DIB, Fallback, Merge, SIB, and Roaring Bitmaps. DIB consistently outperforms the other algorithms across all set sizes, with a more pronounced advantage for larger sets.
    }
\end{figure*}

We measure the time to perform $10^7$ verifications using \dib{}, \fallback{}, \merge{}, \sibt{}, and \roaring{} over the set size (length-partitioned dataset, cf.\ Section~\ref{sec:exp:setup:datasets}) for $\winklerTau{\in}\{0.85, 0.9, 0.95\}$. 
Figure~\ref{fig:overlap_verification} shows the result for $\winklerTau{=}0.9$, the other results are similar. 

On average across all thresholds and string lengths, \dib{} is faster than \fallback{}, \merge{}, \roaring{}, and \sibt{} by factors of $1.7$, $9.0$, $9.6$, and $2.3$, respectively. The advantage over other bitmap techniques stems from differing design goals. \fallback{}, \sibt{}, and \roaring{} compute the exact overlap of two sets to determine if a pair meets a threshold. In contrast, \dib{} uses compressed, approximate bitmaps that correctly solve the decision problem for $98.7\%$ of the pairs and resorts to the slower exact overlap computation (\fallback{}) only for the remaining $1.3\%$.

\subsection{Evaluation of Jaro-Winkler Verification}
\label{sec:exp:jw-verification}

\begin{figure*}[h]
    \centering
    \includegraphics[width=\figscale\textwidth]{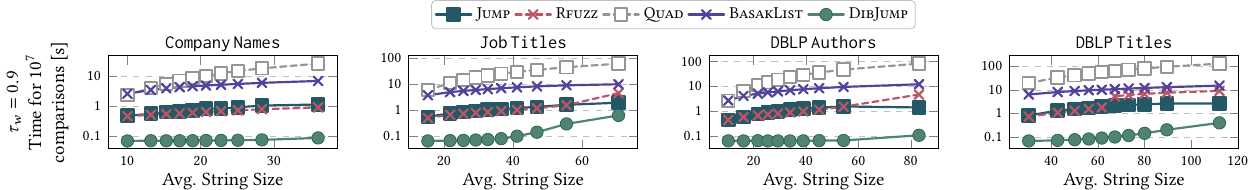} 
    \caption{Runtime comparison of different Jaro-Winkler verification algorithms.}
    \label{fig:verification}
    \Description{
        The line plot shows the runtime comparison of different Jaro-Winkler verification algorithms for 10 million verifications at a Jaro-Winkler threshold of 0.9. The x-axis shows the set size, while the y-axis shows the runtime in seconds per 10 million comparisons. The algorithms compared are JUMP, QUAD, RapidFuzz, BasakList and DIBJUMP. DIBJUMP consistently outperforms the other algorithms across all set sizes due to DIBs filtering. When comparing the non-filtered algorithms, JUMP and RapidFuzz perform similarly for small strings, while JUMP outperforms RapidFuzz for longer strings. QUAD is the slowest algorithm and BasakList is faster than QUAD but slower than JUMP and RapidFuzz.
    }
\end{figure*}

We evaluate the performance of \linearver{} against \quadver{}, \rapidfuzz{}, \relwork{and \basakunopt{}} over the set size (length-partitioned dataset, cf.\ Section~\ref{sec:exp:setup:datasets}) for $\winklerTau\in\{0.85, 0.9, 0.95\}$. We also consider \diblinver{}, the combined verification technique used in our solution that filters pre-candidates using \dib{} before evaluating \linearver{}. Figure~\ref{fig:verification} shows the average time to perform $10^7$ verifications at $\winklerTau{=}0.9$; the other results are similar. We report averages across all threshold values.

Expectedly, \quadver{} is the slowest algorithm.
\linearver{} and \rapidfuzz{} perform similarly for small strings (${\leq}64$, $\pm 6\%$ on average), with \linearver{} outperforming \rapidfuzz{} by $3.2{\times}$ on average for longer strings (${>}64$).
This is due to \rapidfuzz{}'s inherently quadratic runtime, which is accelerated using bitwise operations: 
strings longer than $64$ characters do not fit into a single register and require a quadratic number of bitmap comparisons.
In contrast, \linearver{} runs in linear time and scales beyond 64 characters.
\relwork{\basakunopt's performance is between \linearver{}/\rapidfuzz{} and \quadver{}: On average, \linearver{} outperforms \basakunopt{} by $6.9\times$.
    While \basakunopt{} runs in linear time like \linearver{}, \basakunopt{} does not utilize the positional filter to prune pairs early (cf.\ Section~\ref{sec:jaro-winkler-similarity-computation}) and requires an additional indirection, making it less cache-friendly.
    The positional filter is particularly effective and increases \linearver{}'s performance by $2.4\times$; adding frequency-based sorting of characters increases this further to a factor of $2.6$.}

In \diblinver{}, we combine \dib{} filtering with \linearver{}.
On average, only $2.7\%$ of all verifications require \linearver{}; other verifications are filtered by \dib{}.
As \dib{} filtering is highly efficient, \diblinver{} outperforms all other non-filtered verification methods by 1--3 orders of magnitude (cf.\ Section~\ref{sec:exp:dib-vs-ovlp}), with an average speedup of $8$, $10.7$, \relwork{$55.9$}, and $203$ over \linearver{}, \rapidfuzz{}, \relwork{\basakunopt{}}, and \quadver{}, respectively.

\subsection{Evaluation of Jaro-Winkler Similarity Join}
\label{sec:exp:jw-join}

\begin{figure*}[h]
    \centering
    \includegraphics[width=\figscale\textwidth]{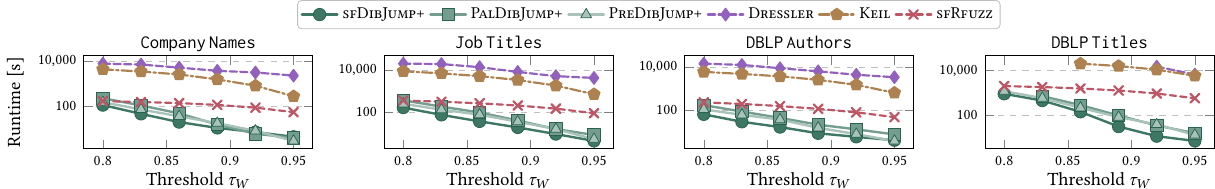}
    \caption{Runtime comparison of different Jaro-Winkler similarity join algorithms for $10^5$ strings.}
    \label{fig:algos-100k}
    \Description{
        The line plot shows the runtime comparison of different Jaro-Winkler similarity join algorithms for 100,000 strings. The x-axis shows the Jaro-Winkler thresholds, while the y-axis shows the runtime in seconds. The algorithms compared are SF-DIBJUMP+, PAL-DIBJUMP+, PRE-DIBJUMP+, Dressler, Keil, and SF-RapidFuzz. Dressler, Keil and SF-RapidFuzz are significantly slower than the other algorithms. The plot illustrates the effectiveness of the proposed algorithms in outperforming the state-of-the-art methods over multiple thresholds and datasets.
    }
\end{figure*}

We evaluate the performance of end-to-end Jaro-Winkler similarity joins. Our solutions \sfdiblinverp{}, \paldiblinverp{}, and \prediblinverp{} all use the \puma{} framework, our \dib{} filter, and \linearver{}, our linear-time verification; they differ only in the pre-candidate generation technique (\sizefilter{}, \prefix{}, and \palloc{}). We compare against the state-of-the-art algorithms \dressler~\cite{Dressler:OnEfficientExecutionBoundedJaroWinklerDistances} and \keil~\cite{Keil:EfficientBoundedJaroWinklerSimilarityBasedSearch}. The baseline \sfrfuzz{} applies size filtering and verifies pre-candidates using an optimized implementation~\cite{RapidFuzz:2025} of the standard Jaro-Winkler verification algorithm. Figure~\ref{fig:algos-100k} shows the runtime of all methods over Jaro-Winkler thresholds $\winklerTau$ for all datasets. Due to their high runtimes, we evaluate \dressler{} and \keil{} only on a sample of size $10^5$.

On average across all datasets and thresholds, \sfdiblinverp{} is the fastest algorithm, outperforming \prediblinverp{}, \paldiblinverp{}, \dressler{}, \keil{}, and \sfrfuzz{} by factors of $1.7$, $1.9$, $612$, $260$, and $16$, respectively.
The large performance gap to the state-of-the-art algorithms \dressler{} and \keil{} has two main causes:
\begin{enumerate*}
    \item \dressler{} and \keil{} are trie-based, whereas \sfdiblinverp{}, \prediblinverp{}, and \paldiblinverp{} either use efficient inverted lists or no indexes.
          \dressler{} and \keil{} incur expensive branch-and-bound exploration such that large parts of the trie must be accessed.
          In particular, \dressler{} essentially performs overlap-based filtering (cf.\ Section~\ref{sec:preliminaries}) using the trie. 
          This is less selective than \keil{}, which computes bounds directly on the Jaro-Winkler similarity and explores a smaller subregion of the trie.
    \item \dressler{} and \keil{} use quadratic verification, either explicitly (\dressler{}), or implicitly during trie traversal (\keil{}).
          This significantly increases the verification cost (cf.\ Section~\ref{sec:exp:jw-verification}).
\end{enumerate*}
The \texttt{DBLP\,Titles} dataset has the longest strings on average (cf.\ Table~\ref{tab:datasets}) and is particularly affected due to deep trie structures and expensive quadratic verification. For this dataset, \sfdiblinverp{} outperforms \dressler{} and \keil{} by factors of $1131$ and $615$, respectively.

\begin{figure*}[h]
    \centering
    \includegraphics[width=\figscale\textwidth]{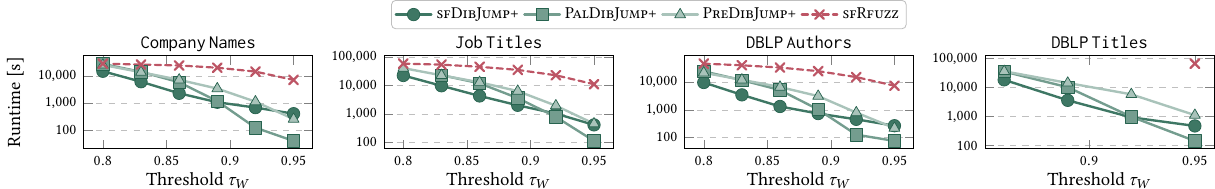}
    \caption{Runtime comparison of different Jaro-Winkler similarity join algorithms for $10^6$ strings.}
    \label{fig:runtime-jw-simjoin}
    \Description{
        The line plot shows the runtime comparison of different Jaro-Winkler similarity join algorithms for 1 million strings. The x-axis shows the Jaro-Winkler thresholds, while the y-axis shows the runtime in seconds. The algorithms compared are SF-DIBJUMP+, PAL-DIBJUMP+, PRE-DIBJUMP+, and SF-RapidFuzz. SF-DIBJUMP+ outperforms the other algorithms for low thresholds, while PAL-DIBJUMP+ performs better for high thresholds. SF-RapidFuzz is significantly slower than the other algorithms across all thresholds.
    }
\end{figure*}

Hence, we limit further discussion to \sfdiblinverp{}, \prediblinverp{}, \paldiblinverp{}, and \sfrfuzz{}. Figure~\ref{fig:runtime-jw-simjoin} shows their performance on larger samples of size $10^6$.
Due to \sfrfuzz{}'s high runtimes, we increase its timeout to 24 hours.

We first discuss \sfrfuzz{}.
On average, \sfdiblinverp{} outperforms \sfrfuzz{} by a factor of $48$. \sfdiblinverp{} and \sfrfuzz{} differ in three aspects:
\begin{enumerate*}
    \item overlap computation,
    \item Jaro-Winkler verification, and
    \item \puma{} usage.
\end{enumerate*}
\puma{} affects the pre-candidate generation \sizefilter{} used by both algorithms: using \puma{}, a tighter bound for $\jaroTau$ (cf.\ Section~\ref{sec:winkler-trie} and Lemma~\ref{lem:size-filter}) results in a tighter size filter. This results in an average reduction of pre-candidates by $36\%$ for \sfdiblinverp{} compared to \sfrfuzz{}. For \sfdiblinverp{}, these pre-candidates are then efficiently and effectively filtered using \dib{} before verification with \linearver{}. \sfrfuzz{}, in contrast, directly verifies the Jaro-Winkler similarity. As discussed in Section~\ref{sec:exp:jw-verification}, \diblinver{} is $10.6{\times}$ faster than \rapidfuzz{} on average across string lengths.\footnote{In the similarity join setting, the verified pairs differ from the samples used in Section~\ref{sec:exp:jw-verification}. Furthermore, \dib{}'s cost model is optimized for the similarity join setting; hence, \sfdiblinverp{} performs better than predicted from the individual components.}

\review[3]{We now discuss \sfdiblinverp{}, \prediblinverp{}, and \paldiblinverp{}. Cost-model based optimization of \dib{} is only a minor part of the overall runtime (${<}1.2\%$ on average, cf.\ Section~\ref{sec:optimal-bitmap-sizes})}. Across all datasets, \sfdiblinverp{} outperforms \prediblinverp{} and \paldiblinverp{} by factors of $2.7$ and $1.6$, respectively, but the performance depends on the threshold $\winklerTau{}$. \sfdiblinverp{} performs well at low thresholds (${<}0.9$), while \paldiblinverp{} performs well at high thresholds (${>}0.9$). Although \prefix{} and \palloc{} never produce more pre-candidates than \sizefilter{}, even for low thresholds, they incur more index reads, i.e., elements accessed (in the inverted lists) to generate pre-candidates. In particular, they generate only $57\%$ and $47\%$ of \sizefilter{}'s pre-candidates on average, but require $159\%$ and $208\%$ of its index reads, respectively. This results in higher runtimes ($265\%$ and $163\%$ of \sfdiblinverp{}). For high thresholds (${>}0.9$), in contrast, \palloc{} generates $0.047\%$ of \sizefilter{}'s pre-candidates with $0.081\%$ of its index reads. For these thresholds, \paldiblinverp{} outperforms \sfdiblinverp{} by $2.5\times$ on average. Due to the low universe size $|\setuniverse|$, \prediblinverp{} does not perform as well as \sfdiblinverp{} and \paldiblinverp{}. \review[2]{Automated selection of a pre-candidate generation method is orthogonal to our work; existing meta-algorithms like \textsc{Fast}~\cite{Schmitt:ExtensibleRobustSimilarityQueries} select the best pre-candidate generation method for a given dataset and threshold at runtime.} \review[1]{We additionally compare \sfdiblinverp{} to \sfdibrfuzzp{} that replaces our verification method \linearver{} with \rapidfuzz{}. On average, both methods perform the same (${<}{\pm}1\%$). \rapidfuzz{} tends to slightly outperform \linearver{} on small strings while \linearver{} scales better to large strings (cf.\ Section~\ref{sec:exp:jw-verification}). Similarly, \sfdiblinverp{} is faster than \sfdibrfuzzp{} on datasets with large strings (\texttt{DBLP\,Titles}, 15\% faster), and slower on datasets with small strings (\texttt{Job\,Titles}, 13\% slower).}

\section{Related Work} \label{sec:related-work}

\paragraph{\textbf{Jaro-Winkler Similarity Joins}}

Dressler et al.~\cite{Dressler:TimeEfficientExecutionBoundedJaroWinklerDistances} propose overlap and size filters to reduce the number of comparisons for the Jaro-Winkler similarity join.
They store sorted strings (w.r.t.\ a global character order) in a standard trie that encodes the prefix of the sorted strings with the full strings only stored at the leafs.
The trie is traversed in character order and strings satisfying the overlap filter are pre-candidates, which are verified w.r.t.\ the Jaro-Winkler similarity.
This method is effective for very high similarity thresholds, but shows poor performance for thresholds ${\leq} 0.85$~\cite{Dressler:OnEfficientExecutionBoundedJaroWinklerDistances}.

Keil et al.~\cite{Keil:EfficientBoundedJaroWinklerSimilarityBasedSearch} build a Patricia tree over the input strings and apply branch-and-bound to traverse the tree and compute upper bounds on the Jaro-Winkler similarity.
While this allows entire subtrees to be skipped, multiple trie traversals are required for different string lengths.
The approach was evaluated only for high thresholds, small query sets, and large target sets; our experiments suggest poor performance for similarly sized query and target sets.

Wang et al.~\cite{Wang:EfficientApproximateEntityMatching} present a filter-and-verify approach that uses an index based on special string signatures.
Since Keil et al.~\cite{Keil:EfficientBoundedJaroWinklerSimilarityBasedSearch} already identified shortcomings in this paper that render it impossible to reimplement it, we exclude it from our experimental evaluation.

Overall, none of these approaches scale well to large datasets, low thresholds, or small alphabets.

\relwork{%
    In addition, we discuss algorithms to verify candidate pairs w.r.t.\ the Jaro-Winkler similarity.
    Basak et al.~\cite{Basak:FastAlgorithmsComputingJaroSimilarity,Basak:OnComputingJaroSimilarity} compute the Jaro and the Jaro-Winkler similarity of two strings of max.\ length $n$ over an alphabet $\Sigma$ in $\mathcal{O}(n + |\Sigma|)$ time and space.
    Their practically optimized implementation requires $\mathcal{O}(|\Sigma|\cdot n)$ space.
    During preprocessing, they count character occurrences and construct lists of the positions of each character for both strings.
    These lists are organized in array of size $|\Sigma|$, where each entry points to the corresponding list.
    During verification, the position lists of each occurring character are merged to determine matches and their positions, which are then used to compute transpositions and the Jaro-Winkler similarity.
    
    In contrast, our sort-based \linearver algorithm avoids per-character lists, achieving $\mathcal{O}(\min\{n + |\Sigma|, n\log n\})$ time and $\mathcal{O}(n + |\Sigma|)$ space when $|\Sigma| = o(n\log n)$, and $\mathcal{O}(n)$ space otherwise.
    Our experiments show that our algorithm with the positional filter outperforms Basak et al.~\cite{Basak:FastAlgorithmsComputingJaroSimilarity,Basak:OnComputingJaroSimilarity} by $6.9{\times}$ on average across all instances. %
}

\paragraph{\textbf{Set Similarity Joins}}

State-of-the-art set similarity join techniques rely on signature-based frameworks, i.e., (1) variations of the \textit{prefix filter}~\cite{Chaudhuri:PrimitiveOperatorSimilarityJoins, Mann:EmpiricalEvaluationSetSimilarityJoin} or (2) the \textit{partition-and-enumeration framework}~\cite{Arasu:EfficientExactSetSimilarityJoins} with optimizations like asymmetric enumeration~\cite{Norouzi:FastSearchHammingSpaceMultiIndexHashing}, dimension rearrangement~\cite{Zhang:HmSearchEfficientHammingDistanceQueryProcessingAlgorithm}, and cost-based partition skipping~\cite{Deng:EfficientPartitionBasedMethodExactSetSimilarityJoinsPARTALLOC, Qin:GeneralizingPigeonholePrincipleSimilaritySearchHammingSpace, Liu:HAPHammingSpaceIndexAugmentedPigeonholePrinciple}.

Recent works improve the performance by grouping index entries and probing signatures~\cite{Wang:LeveragingSetRelationsExactDynamicSetSimilarityJoin} or by processing sets differently based on their size~\cite{Deng:OverlapSetSimilarityJoinsTheoreticalGuaranteesSIZEAWARE}.
Moreover, multiple algorithms have been combined to achieve lower runtimes~\cite{Deng:OverlapSetSimilarityJoinsTheoreticalGuaranteesSIZEAWARE} and robustness~\cite{Schmitt:TwoLevelSignatureSchemeStableSetSimilarityJoinsTWOL,Widmoser:MetricJoinLeveragingMetricPropertiesRobustExactSetSimilarityJoins,Schmitt:ExtensibleRobustSimilarityQueries}.

Sandes et al.~\cite{Sandes:BitmapFilterSpeedingUpExactSetSimilarityJoinsBitwiseOperations} use fixed-size hashed bitmaps to derive an upper bound on the overlap similarity and prune the search space.
While a \dib{}'s hashed part resembles this, a \dib{}'s exact part additionally provides an effective lower bound to directly forward pairs to the result.
Section~\ref{sec:exp:dib-vs-ovlp} shows that \dib{}s outperform this approach.

All known set similarity join algorithms are orthogonal and can be integrated into PUMA to benefit from our efficient \dib{} structure.

\paragraph{\textbf{Efficient Overlap Computation}}

Existing set intersection algorithms often use merge-based approaches~\cite{Ding:FastSetIntersectionMemory} on two sorted sets $r$, $s$ that are scanned side by side, requiring only $\mathcal{O}(|r| + |s|)$ time.
Section~\ref{sec:exp:dib-vs-ovlp} shows that our \dib{}s outperform this merge-based method.

Demaine et al.~\cite{Demaine:AdaptiveSetIntersectionsGALLOPING} propose \emph{Galloping} and use binary search to find similar elements in two sorted sets $r$, $s$.
For each element in $r$, they search for the corresponding element in $s$ in $\mathcal{O}(\log |s|)$ time, requiring $\mathcal{O}(|r| \log |s|)$ time overall.
While Galloping skips many comparisons if $|r| \ll |s|$, we exclude it from our experiments as the merge-based method typically outperforms it if $|r| \approx |s|$.

Graph set intersection is a special form of set intersection.
Zheng et al.~\cite{Zheng:AcceleratingSetIntersectionsGraphsReducingMerging} present \emph{RCode}, which enhances graph set intersection using a vertex labelling technique.
Zhang et al.~\cite{Zhang:BigSetEfficientSetIntersectionApproach} revisit RCode and show inefficiencies for the general case. %
Hence, they introduce \emph{BigSet}, which partitions elements into buckets, using bitwise operations to find candidate buckets.
While fast for large sets, the speedup on small sets is limited to $2\times$ compared to merge-based methods.
Since general set intersection is applicable to graphs but not necessarily vice versa, we exclude them from evaluation.

Yang et al.~\cite{Yang:HEROHierarchicalSetPartitioningJoinFrameworkSpeedingUpSetIntersectionOverGraphs} propose \emph{SIB trees} and the \emph{HERO} framework to speed up set intersection on graph data using hierarchical partitioning.
Each set is recursively segmented into (non-empty) partitions and the overlap is computed by incrementally intersecting the partition identifiers until reaching the base set elements.
Their approach minimizes redundant comparisons, scales well to large graphs, and offers significant performance improvements over standard set intersection methods.
SIB trees can be effectively used for general set intersection, and we compare against them in Section~\ref{sec:exp:dib-vs-ovlp}.

Finally, SIMD has been used to accelerate overlap computations.
Han et al.~\cite{Han:SpeedingUpSetIntersectionsGraphAlgorithmsSIMD} improve the merge-based method using SIMD and a byte-checking filter.
Zhang et al.~\cite{Zhang:FESIAFastSIMDEfficientSetIntersection} use fixed-size bitmaps and a kernel set intersection technique to matching elements.
While SIMD is an orthogonal optimization, \dib{}s are independent of SIMD.
Hence, we do not compare against SIMD-powered approaches.

\section{Conclusion} \label{sec:conclusion}

We proposed a novel, scalable approach for efficient Jaro-Winkler similarity joins with three key contributions:
(1) The \puma{} framework groups strings by prefix to tighten filter conditions and reduce the number of similarity computations.
(2) Dual-bound intersection bitmaps (\dib{}s) encode strings into fixed-size bitmaps with an exact and a hashed part, providing effective overlap bounds; 
a cost model adapts \dib{}s to the input data.
\relwork{(3) \linearver{} computes the Jaro-Winkler similarity in linear time for bounded alphabet sizes and scales to long strings; unlike existing linear methods, \linearver\ is also fast for short strings.
}
In our experiments, we outperform the state of the art by up to three orders of magnitude.

\begin{acks}
    This research was funded in whole or in part by the \grantsponsor{FWF}{Austrian Science Fund (FWF)}{https://www.fwf.ac.at/} \grantnum{FWF}{10.55776/P34962}, by the \grantsponsor{Salzburg}{State of Salzburg, Austria}{https://www.salzburg.gv.at/}: ref. no.~\grantnum{Salzburg}{20204-WISS/262/9-2021}, by \grantsponsor{Celonis}{Celonis}{} through the project \emph{SPRING -- Scalable Process Mining}, and by the Database Research Cluster at the University of Salzburg, Core Facility (CF)~3914 of the Austrian Federal Ministry for Women, Science, and Research (\url{https://forschungsinfrastruktur.bmfwf.gv.at/en/fi/_3914}).
\end{acks}

\bibliographystyle{ACM-Reference-Format}
\bibliography{bibliography}
\newpage
\clearpage

\appendix

\section{Quadratic-Time Jaro-Winkler Verification}
The original algorithm for Jaro-Winkler verification scans the two strings in a greedy manner to find matches, i.e., pairs of characters that are identical and satisfy the window constraint.

\begin{algorithm}[htbp]
    \small
    \DontPrintSemicolon
    \KwIn{Strings \(r, s\).}
    \KwOut{Jaro similarity score \(S_J(r,s)\).}
    \SetKwFunction{JaroGreedy}{JaroGreedy}
    \SetKwProg{Fn}{Function}{:}{}

    \Fn{\JaroGreedy{\(r, s\)}}{
        \( \omega \gets \lfloor \max\{|r|, |s|\}/2 \rfloor - 1 \)\;
        \( I_M \gets \emptyset \), \( J_M \gets \emptyset \) \AlgoComment{Sets of matched indices}

        \AlgoComment{Greedy Matching Phase}
        \For{\( i \gets 1 \) \KwTo \( |r| \)}{ \label{line:greedy_jaro_outer_loop}
            \( \textit{start} \gets \max\{1, i - \omega\} \), \( \textit{end} \gets \min\{|s|, i + \omega\} \)\;
            \For{\( j \gets \textit{start} \) \KwTo \( \textit{end} \)}{
                \If{\( r[i] = s[j] \) \textbf{and} \( j \notin J_M \)}{
                    \( I_M \gets I_M \cup \{i\} \)\;
                    \( J_M \gets J_M \cup \{j\} \)\;
                    \textbf{break} \AlgoComment{Greedily take first match}
                }
            }
        }

        \( m \gets |I_M| \)\;
        \lIf{\( m = 0 \)}{\Return 0}

        \AlgoComment{Transposition Computation}
        Let \( \hat{r} \) be the substring of \( r[i] \) for all \( i \in I_M \) (sorted by \(i\))\;
        Let \( \hat{s} \) be the substring of \( s[j] \) for all \( j \in J_M \) (sorted by \(j\))\;
        \( t \gets \frac{\hat{r} \Delta \hat{s}}{2} \)\;

        \Return \( \frac{1}{3} \left( \frac{m}{|r|} + \frac{m}{|s|} + \frac{m - t}{m} \right) \)\;
    }
    \caption{Greedy Jaro Similarity}
    \label{alg:greedy_jaro}
\end{algorithm}

\begin{figure*}[t]
    \centering
    \includegraphics[width=0.9\textwidth]{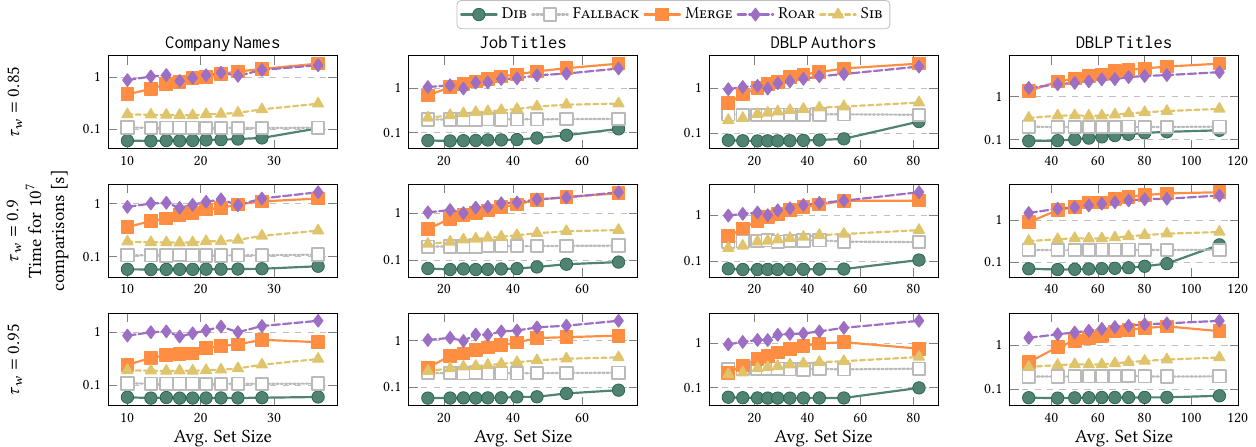}
    \caption{Runtime comparison of different overlap verification algorithms.}
    \label{fig:overlap_verification_full}
    
    \vspace{2em} %
    
    \includegraphics[width=0.9\textwidth]{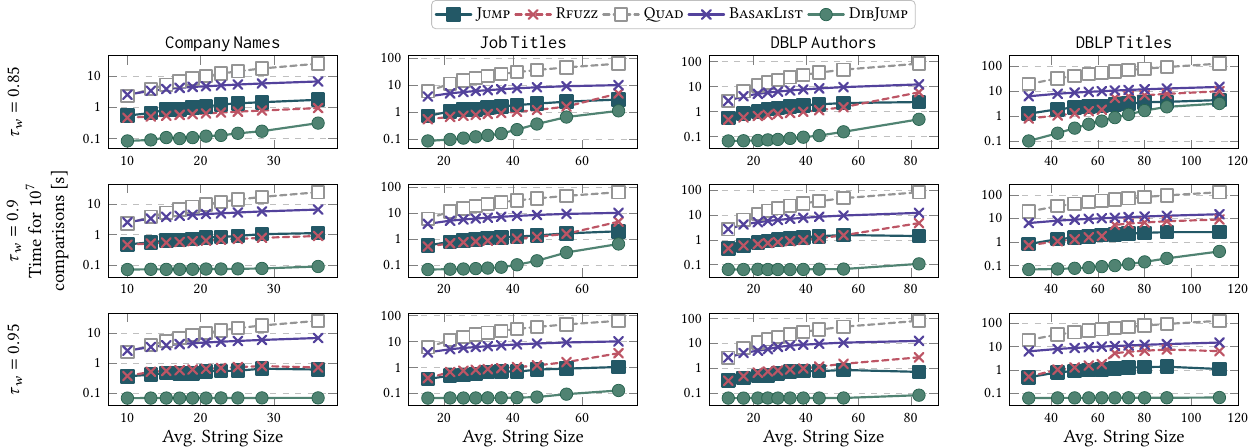}
    \caption{Runtime comparison of different Jaro-Winkler verification algorithms.}
    \label{fig:verification_full}
\end{figure*}
\section{Proofs} \label{sec:proofs}

\paragraph{\textbf{Notation}}

$r$, $s$ denote sets and $B_r$, $B_s$ the corresponding DIBs. $\land$ and $\oplus$ denote the bitwise $\mathtt{AND}$ and $\mathtt{XOR}$, respectively, and $\pc{B_r}$ counts the number of set bits in $B_r$.

For any mapping $M\subseteq[|r|]\times[|s|]$ we write
$I_M=\{\,i\mid \exists j:\ (i,j)\in M\,\}$ and
$J_M=\{\,j\mid \exists i:\ (i,j)\in M\,\}$.
Equivalently, for a specific matching $G$ we write $I_G,J_G$.

\begin{cleverproof}{thm:correctness_lemur}
    Algorithm~\ref{alg:verifyJW} explicitly enforces character identity (line~\ref{alg:verifyJW:charidentity}) and the window constraint (line~\ref{alg:verifyJW:matchingcase1-start}), ensuring validity.
    In addition, occurrences of any given character are considered in increasing order of their positions in both strings.
    
    We prove maximality of the matching by induction on the number of matches found by the algorithm.
    Let $M^{\ast}$ be an optimal valid matching and let $M$ be the matching constructed by Algorithm~\ref{alg:verifyJW}.
    Define $M_k = \{(i,j)_l \in M \mid l \leq k\}$ as the first $k$ matches found by the algorithm.
    We claim that for every $k \geq 0$, there exists a valid matching $M'$ with $|M'| = |M^{\ast}|$ and $M_k \subseteq M'$.
    We prove the claim by induction on $k$.
    
    For the base case $k=0$, we have $M_0=\emptyset$ and thus the claim holds trivially.
    For the inductive step, assume the claim holds for $k$.
    Let $M'$ be a valid matching with $|M'| = |M^{\ast}|$ and $M_k \subseteq M'$.
    Let $(i,j)_{k+1}$ be the next match found by the algorithm.
    Let $M_k^c = M' \setminus M_k$ be the remaining matches in $M'$ after removing the matches from $M_k$.
    By the inductive hypothesis, $|M_k^c| = |M'| - |M_k| = |M^{\ast}| - |M_k|$.
    We need to show that there exists a valid matching $M''$ with $|M''| = |M^{\ast}|$ and $M_{k+1} \subseteq M''$.
    If $(i,j)_{k+1} \in M_k^c$, we can set $M'' = M'$ and the claim holds.
    Otherwise, since $(i,j)_{k+1}$ is a valid match, at least one of $i$ or $j$ must be matched in $M_k^c$ (otherwise we could add $(i,j)_{k+1}$ to $M'$ to obtain a valid matching larger than $M^{\ast}$, contradicting the maximality of $M^{\ast}$).
    Since the algorithm considers indices in increasing order, any conflicting match $(i,j') \in M_k^c$ implies $j' \succ j$, and any $(i',j) \in M_k^c$ implies $i' \succ i$.
    There are two cases:
    (1) One of $i$ or $j$ is matched in $M_k^c$, but not both.
    W.l.o.g., assume that $j$ is unmatched in $M_k^c$, then $i$ must be matched to some $j'$. We can construct $M'' = M_k \cup (M_k^c \setminus \{(i,j')\}) \cup \{(i,j)_{k+1}\}$.
    (2) Otherwise, there exists some $(i',j) \in M_k^c$ with $i' \neq i$.
    If $(i,j') \in M_k^c$, then we can construct $M'' = M_k \cup (M_k^c \setminus \{(i,j'),(i',j)\}) \cup \{(i,j)_{k+1},(i',j')\}$.
    Rewriting the window constraints as inequalities immediately shows that $(i',j')$ is a valid match.
    In all cases, $M''$ is a valid matching with $|M''| = |M^{\ast}|$ and $M_{k+1} \subseteq M''$, completing the inductive step.
    Thus, by induction, the algorithm constructs a maximal valid matching.
    
    To show minimal cost, observe that the algorithm always considers indices in increasing order.
    When a match is found (line~\ref{alg:verifyJW:charidentity}), it is formed between the smallest available indices $i$ and $j$.
    This local greedy choice minimizes the sum $i+j$, satisfying the minimal cost condition.
    Hence, algorithm~\ref{alg:verifyJW} constructs a Jaro mapping.
\end{cleverproof}

\begin{proof}[Proof of Lemma~\ref{lem:unique-jaro-mapping}]
    We prove the Uniqueness of $I_M$ and $J_M$ for a given Jaro Mapping $M$ by contradiction.
    Assume that there exist two Jaro mappings $M_1\neq M_2$ with $I_{M_1}\neq I_{M_2}$ (the case $J_{M_1}\neq J_{M_2}$ is analogous). 
    By Definition~\ref{def:jaro-mapping} we have $|M_1|=|M_2|$ and both matchings achieve the same (minimal) cost $C=\sum_{(i,j)\in M} (i+j)$.
    
    Consider the symmetric difference $M_1\Delta M_2=(M_1\setminus M_2)\cup(M_2\setminus M_1)$.
    The edges of $M_1\Delta M_2$ form a vertex-disjoint union of alternating paths and cycles (edges alternate between $M_1$ and $M_2$), since each vertex has degree at most 2.
    Let $H$ be the connected component of $M_1\Delta M_2$ that contains the smallest row-index $i^\ast$ from $I_{M_1}\Delta I_{M_2}$.
    Because $|M_1|=|M_2|$, an alternating path must contain an equal number of edges from $M_1$ and $M_2$.
    By construction $H$ is an alternating path that starts with an $M_1$-edge incident to $i^\ast$ and ends with an $M_2$-edge incident to some $i_k\in I_{M_2}\setminus I_{M_1}$.
    Write the edges of $H$ in order as
    \[
        \underbrace{(i^\ast,j_1)}_{\in M_1},\ \underbrace{(i_1,j_1)}_{\in M_2},\ \underbrace{(i_1,j_2)}_{\in M_1},\ \underbrace{(i_2,j_2)}_{\in M_2},\ \dots,\ \underbrace{(i_{k-1},j_k)}_{\in M_1},\ \underbrace{(i_k,j_k)}_{\in M_2}.
    \]
    
    Since $M_1$ and $M_2$ are both minimum-cost matchings, the cost of $M_1 \cap H$ and $M_2 \cap H$ must be equal.
    If this were not true, say $M_1\cap H$ has lower cost than $M_2\cap H$, we could replace $M_2\cap H$ by $M_1\cap H$ in $M_2$ to obtain a matching of lower cost than $M_2$, contradicting the minimality of $M_2$.  
    Therefore we may restrict the cost equality to $H$:
    \[
        (i^\ast+j_1)+(i_1+j_2)+\cdots+(i_{k-1}+j_k)
        \;=\;
        (i_1+j_1)+(i_2+j_2)+\cdots+(i_k+j_k).
    \]
    Subtracting common terms yields $i^\ast = i_k$.
    
    But $i^\ast\in I_{M_1}\setminus I_{M_2}$ while $i_k\in I_{M_2}\setminus I_{M_1}$, so $i^\ast\neq i_k$, a contradiction.
    Therefore $I_{M_1}=I_{M_2}$, and by the same argument $J_{M_1}=J_{M_2}$.
\end{proof}

\begin{cleverproof}{thm:original-greedy-jaro-mapping}
    Trivially, the original greedy algorithm produces a valid mapping with window size $\omega :=\lfloor \frac{\max\{|r|, |s|\}}{2} \rfloor - 1$. 
    Further, the cost of the produced mapping is minimal since the algorithm always matches the first possible pair of characters. 
    The maximality of the produced mapping follows from the same argument as in the proof of Theorem~\ref{thm:correctness_lemur} but by induction on the running variable $i$ of the outer loop of the original greedy algorithm (Line~\ref{line:greedy_jaro_outer_loop}).
    
    Thus, the algorithm produces a Jaro mapping.
\end{cleverproof}

\begin{proof}[Proof of Lemma~\ref{lem:dib-bounds}]
    Let $r$ and $s$ be partitioned into exact and hashed subsets, i.e., $r = r_E \cup r_H$ and $s = s_E \cup s_H$. 
    Since the subsets are disjoint, we have $|r| = |r_E| + |r_H|$ and $|s| = |s_E| + |s_H|$ and also $|r \cap s| = |r_E \cap s_E| + |r_H \cap s_H|$.
    
    In the exact parts $\exactsimple{r}$ and $\exactsimple{s}$, each bit directly encodes a unique token.
    Hence, the bitwise $\mathtt{AND}$ between those parts corresponds to the exact overlap:
    \begin{align}
        |r_E \cap s_E| = \pc{\exactsimple{r} \land \exactsimple{s}} = \varphi_l \label{eq:proof-dib-exact}
    \end{align}
    This provides our lower bound $\varphi_l \leq |r \cap s|$. 
    
    In the hashed parts $\heuristicsimple{r}$ and $\heuristicsimple{s}$, multiple tokens may be mapped to the same bit. Following Eq.~\ref{eq:bitmap-upper-bound-Xor}, we have:
    \begin{align}
        |r_H \cap s_H| \leq \left\lfloor \frac{|r_H| + |s_H| - \pc{\heuristicsimple{r} \oplus \heuristicsimple{s}}}{2} \right\rfloor \label{eq:proof-dib-hashed}
    \end{align}
    
    The total intersection is upper-bounded by summing Equations \ref{eq:proof-dib-exact} and \ref{eq:proof-dib-hashed}, directly yielding the upper bound $\varphi_u$ proposed in lemma~\ref{lem:dib-bounds}:
    \begin{align}
        |r \cap s| \leq \varphi_l + \left\lfloor \frac{|r_H| + |s_H| - \pc{\heuristicsimple{r} \oplus \heuristicsimple{s}}}{2} \right\rfloor
    \end{align}
    
\end{proof}

\section{Further Experimental Results} \label{sec:appendix-experimental-evaluation}
This section includes further experimental results omitted in the main text. Figure~\ref{fig:overlap_verification_full} shows the time to perform $10^7$ verifications using \dib{}, \fallback{}, \merge{}, \sibt{}, and \roaring{} over the set size (length-partitioned dataset, cf.\ Section~\ref{sec:exp:setup:datasets}) for $\winklerTau{\in}\{0.85, 0.9, 0.95\}$.

Figure~\ref{fig:verification_full} shows the average time of \linearver{}, \quadver{}, \rapidfuzz{}, \basakunopt{}, and \diblinver{} to perform $10^7$ verifications for $\winklerTau{\in}\{0.85, 0.9, 0.95\}$.

\end{document}